\documentclass[useAMS,usenatbib]{mn2e}
\usepackage{psfrag,graphicx}
\usepackage{color}
\title[The Impact of Lyman alpha Trapping on the Formation of Primordial Objects]
  {The Impact of Lyman alpha Trapping on the Formation of Primordial Objects}
\author[M.A. Latif et al.]
  {M.A.~Latif$^1$,
  S.~Zaroubi$^1$$^,$$^2$,
  M.~Spaans$^1$ \\
   $^1$ Kapteyn Astronomical Institute, University of Groningen, P.O.~Box 800, 9700 AV Groningen\\
   $^2$ Physics Department, Technion, Haifa 32000, Israel}

\date{Accepted 2010 September 30. Received 2010 September 21; in original form 2010 July 2}

\pagerange{\pageref{firstpage}--\pageref{lastpage}} \pubyear{2009}

\def\LaTeX{L\kern-.36em\raise.3ex\hbox{a}\kern-.15em
    T\kern-.1667em\lower.7ex\hbox{E}\kern-.125emX}

\begin{document}
 \bibliographystyle{mn2e}

\label{firstpage}

\maketitle

\begin{abstract}
{Numerous cosmological simulations have been performed to study the formation of the first objects. We present the results of high resolution 3-D cosmological simulations of primordial objects formation using the adaptive mesh refinement code FLASH by including in an approximate manner the radiative transfer effects of Lyman alpha photons. We compare the results of a Lyman alpha trapping case inside gas clouds with atomic and molecular hydrogen cooling cases.}
{The principal objective of this research is to follow the collapse of a zero metallicity halo with an effective equation of state (that accounts for the trapping) and to explore the fate of a halo in each of the three cases, specifically, the impact of thermodynamics on fragmentation of halos.}
{Our results show that in the case of Lyman alpha trapping, fragmentation is halted and a massive object is formed at the center of a halo. The temperature of the gas remains well above $\rm 10^{4}$ K and the halo is not able to fragment to stellar masses. In the atomic cooling case, gas collapses into one or two massive clumps in contrast to the Lyman alpha trapping case. For the molecular hydrogen cooling case, gas cools efficiently and fragments.}
{The formation of massive primordial objects is thus strongly dependent on the thermodynamics of the gas. A salient feature of our results is that for the formation of massive objects, e.g. intermediate mass black holes, feedback effects are not required to suppress $\rm H_{2}$ cooling, as molecular hydrogen is collisionally dissociated at temperatures higher than $\rm 10^{4}$ K as a consequence of Lyman alpha trapping.}
\end{abstract}


\section{Introduction}

The formation of primordial objects is not yet very well understood. A lot of progress has been made in analytical and numerical approaches but still further investigation is required. The formation of the first objects depends on the fragmentation process and its relation to thermodynamical properties of the gas \citep{2005SSRv..116..625C}. In order to understand their formation, it is crucial to investigate the impact of thermodynamics on primordial halos which define the mass scale of first objects. It is generally believed that first stars are formed in minihalos of $\rm \sim 10^{6} M_{\odot}$ at redshift $\rm z \sim 20-30$ while quasars are formed in massive halos at $\rm z \sim 10$ \citep{2004PASP..116..103B}.

Numerical simulations performed to study the formation and fragmentation of primordial gas clouds \citep{1997ApJ...474....1T, 2000ApJ...540...39A,2002Sci...295...93A,2006ApJ...652....6Y, 2007ApJ...654...66O, 2004NewA....9..353B, 2004ARA&A..42...79B, 2005SSRv..116..625C, 2009Natur.459...49B} show that first stars are formed in $\rm \sim 10^{6} M_{\odot}$ halos if gas is condensed by $\rm H_{2}$ cooling. It has been seen in these simulations that first stars are massive due to inefficient cooling (in the absence of metals and dust) and are formed in isolation. Recent work by \cite{2010arXiv1006.1508C} shows that primordial gas is susceptible to fragmentation even with small amount of turbulence. For metal free gas, the only efficient coolants that can bring down the temperature below $\rm \sim10^{4}$ K are $\rm H_{2}$ and HD. Gas will not fragment to form stars in their absence \citep{2000ApJ...540...39A}. Feedback processes can suppress the formation of molecular hydrogen \citep{2000ApJ...534...11H}.

Cooling due to molecular hydrogen in halos having virial temperatures $\rm >10^{4}$ K can be suppressed through Lyman-Werner radiation emitted by a nearby star forming galaxy. Such halos therefore could be suitable for the growth of black holes \citep{2008MNRAS.391.1961D}. Early fragmentation will not occur when $\rm H_{2}$ cooling is suppressed either due to external or internal sources and will favor the formation of massive objects \citep{2002ApJ...569..558O,2003ApJ...596...34B,2006MNRAS.370..289B}. In the absence of molecular hydrogen cooling, gas can only cool through atomic line radiation. \cite{2008ApJ...682..745W} performed cosmological simulations to study the collapse of protogalactic gas clouds cooled by only atomic lines and found that an object of $\rm 10^{5} M_{\odot}$ is formed in the center of a metal free halo. They also determined that angular momentum transfer played a vital role in the formation of a massive object. \cite{2003ApJ...596...34B} used SPH simulations to study the evolution of metal free gas in the absence of $\rm H_{2}$ cooling. Their results show that collapse is isothermal and fragmentation is not very efficient. As a result one or two massive clumps of $\rm 10^{6} M_{\odot}$ are formed containing $\rm >10$\% of gas. \cite{2009MNRAS.393..858R,2009MNRAS.396..343R} carried out cosmological simulations and followed the dynamic evolution of metal free halos cooled by atomic line radiation and found that isothermal collapse leads to the formation of a self gravitating stable disk at the center of a halo.

Massive black holes can only form in atomic cooling halos \citep{2000ApJ...540...39A} with virial temperatures greater than $\rm 10^{4}$ K. Self-gravitating gas can form black holes through direct collapse if it is cooled down to a small fraction of the virial temperature \citep{2009arXiv0904.4247B}. Various other ways have been suggested for the formation of black holes \citep{2004gimi.confE..14H, 2004ApJ...613...36H,2004Natur.428..724P,2009ApJ...694..302D,1978Obs....98..210R,2008arXiv0803.2862D}. Gas in primordial halos can cool through atomic line radiation and collapses isothermally to form a $\rm 10^{4}-10^{6} M_{\odot}$ object if the temperature remains as high as $\rm 10^{4}$ K \citep{2006ApJ...652..902S}. These authors found that trapping Lyman alpha photons stiffens the equation of state and raises the temperature above $\rm 10^{4}K$. The fraction of molecular hydrogen is almost zero at such higher temperatures.

Trapping Lyman alpha photons increases the Jeans mass and fragmentation is likely to be inhibited. The impact of Lyman alpha pressure on accretion of first stars has been studied by \cite{2008ApJ...681..771M}. They found that it can not effect overall infall except in polar regions of 20-30 $\rm M_{\odot}$ stars. The key assumption in most models for the formation of massive objects through direct collapse is that gas must avoid fragmentation otherwise it might form low mass Pop III stars. Hence, quenching the formation of $\rm H_{2}$ will be necessary. The model by \cite{2006ApJ...652..902S} for the formation of pregalactic black holes requires no explicit mechanisms for the destruction of $\rm H_{2}$ as trapping of Lyman alpha photons keeps the gas $\rm H_{2}$ free, but also see \cite{2010ApJ...712L..69S}. Motivated by their model, we have performed 3-D cosmological simulations that include trapping effects of Lyman alpha photons and compared our results with previous work (atomic and $\rm H_{2}$ cooling cases). The only assumption for this model is that the halo is metal free. The metal distribution is generally inhomogeneous and metal enrichment has not proceeded very far at higher redshifts. Pockets of metal free gas can exist thus until redshift six \citep{2009ApJ...700.1672T}.

The primary aim of this research is to follow the collapse of a (close-to) zero metallicity halo and to explore, whether it fragments into multiple system or forms a massive object with out fragmenting. We carry out high resolution cosmological simulations that have a dynamic range of $\rm \sim2\times10^{5}$ in linear scale and study the impact of thermodynamics on the collapse of primordial gas. We present in this paper three sets of high resolution cosmological simulations with different physics ingredients. In the first case, we add atomic cooling assuming that there is a UV background that suppresses molecular hydrogen formation. In the second set of simulations we include cooling due to molecular hydrogen to see how it affects the process of collapse. In the third suite of simulations, we assess the influence of Lyman alpha photon trapping on the formation of primordial objects. We compare the results of the Lyman alpha trapping case with atomic and molecular hydrogen cooling cases. For the case of Lyman alpha trapping, we assume that no molecular hydrogen formation takes place at temperatures above the collision break-up energy of $\rm H_{2}$. Also, even small amounts of dust, if present, can absorb and re-emit the Lyman alpha photons at much lower frequencies \citep{1999ApJ...518..138H}. So, we assume zero dust abundance.

Our paper is organized as follows. In section 2, we describe the simulation setup and numerical methods used in this paper. In section 3, we discuss different heating and cooling mechanisms that we included in our simulations. Finally, in section 4, we discuss the results obtained with our simulations and present our conclusions.

\section {Simulations}

We are using the code FLASH3 \citep{2000ApJS..131..273F}. FLASH is a module based, Eulerian AMR, parallel simulation code which can solve a broad range of astrophysics problems. The Message-Passing Interface (MPI) library is used  to get portability and scalability on many different parallel systems. It makes use of the PARAMESH library to handle a block-structured adaptive grid, to add higher resolution depending upon user's defined refinement criteria. For hydrodynamic calculations we use the directionally split piecewise-parabolic method \citep[PPM,][]{1984JCoPh..54..174C} and inside shock waves we use the HLLE solver. PPM is favored above other methods due to its higher order accuracy and its ability to sharpen shocks and contact discontinuities as it does not use artificial viscosity. In order to calculate the evolution of dark matter, FLASH uses an N-body particle mesh technique and couples it to the  hydrodynamics. The gravity solver used in FLASH is the oct-tree based multigrid Poisson solver developed by \cite{2008ApJS..176..293R}.

We take a three-dimensional box of 10 Mpc (co-moving) on each side. In order to simulate the evolution of dark matter, we are initializing $\rm 2.6 \times 10^{6}$ particles. We use the COSMICS (grafic) package developed by \cite{1995astro.ph..6070B} to introduce Gaussian random field cosmological initial conditions. As motivated by inflation, the density perturbations are assumed to be isotropic, homogeneous and Gaussian. We use standard parameters from the WMAP 5-year data ($\rm \Omega_{m} =0.2581$, $\rm H_{0}=72 kms^{-1}Mpc^{-1}$, $\rm \Omega_{b}=0.0441$) with a value of $\rm \sigma_{8} = 0.8$. We use periodic boundary conditions both for gravity and hydrodynamics.

We run the COSMICS package to create Gaussian random initial conditions and select the set of initial conditions having the highest density peak. We find the location of the highest density peak and shift it to the center of the box. In the region of 1Mpc around the center of the box, we initialize our simulation with a high level of resolution, which corresponds to $\rm 512^{3}$ grid cells. The rest of box is set to a resolution of $\rm 128^{3}$ cells. We start the simulation at redshifts ranging from z=50-70 for different sets of Gaussian random field initial conditions with higher and lower density peaks halos. We select a massive halo and zoom in with the FLASH refinement criteria. As adding higher levels of uniform refinement within the region of interest becomes a computationally demanding task, we exploit the AMR technique to add dynamic levels of refinement. We add 8 additional dynamic levels of refinement using specific density threshold criteria. This gives us 15 levels of refinement in total. The effective resolution in the region of interest is $\rm \sim2\times10^{5}$ in linear scale (i.e. dynamic range). The virial mass of the system is $\rm \sim 3\times 10^{9}M_{\odot}$ and the virial temperature is $\rm \sim 10^{5}$K ($\rm T_{vir} = 10^{4}\times(M_{vir}/M_{\odot})^{3/2}(1+z_{vir}/10))$. The collapse happens at redshift 10.
In the case of molecular hydrogen cooling, the Jeans mass reduces as gas cools to lower temperatures and collapes to higher densities compared to the other simulations, so we add an additional dynamic refinement level to resolve the Jeans mass (maximum refinement in this is 16 levels). Our refinement criteria are good enough to resolve shock waves and they also ensure the fulfillment of the Truelove criteria stating that the Jeans length should be resolved by at least four cells \citep{1997ApJ...489L.179T}. In this way, we avoid artificial fragmentation caused by density perturbations created during discretization of the grid. The top level hierarchy of refinement levels helps in even greater precision.
\section{Physics included}

\subsection {Cooling}

Cooling is an important process in the formation of structures, like stars and galaxies. It is crucial to understand the physics of cooling in primordial gas in order to study the formation of first objects. Adequate amounts of gas has to gather in dark matter halos to form structures. Cooling becomes efficient when the cooling time $\rm t_{cool}= 3nkT/2\Lambda$ becomes less than the free-fall time $\rm t_{ff}=\sqrt{3\pi/32G\rho}$, where n is the number density, $\rm \rho$ is the mass density and $\rm \Lambda$ is the cooling rate. Under this condition gas cools, collapses and can fragment. Mass scales of fragments are determined by the Jeans mass $\rm M_{J} \approx 10^{6}M_{\odot}(T/10^{4}K)^{3/2}$ (for typical densities of $\rm 10^4 cm^{-3}$) which depends on the thermodynamics of the gas. Gas cannot cool below 8000 K if it is cooled by only atomic line radiation. Consequently, the Jeans mass remains $\rm >10^{5} M_{\odot}$ and fragmentation to lower mass scales is inhibited. In order to induce fragmentation of lower mass scales gas should be cooled down to lower temperatures. Introduction of molecular hydrogen and HD cooling can lower the gas temperature down to $\rm \sim 100$ K and reduces the Jeans mass to $\rm \sim 10^{3} M_{\odot}$. Fragmentation at lower than the Jeans scale is prohibited as gas becomes pressure supported.

The net cooling function of a gas is a combination of processes, like line radiation, photo-ionization, compton scattering, bremsstrahlung, radiative recombination, etc. The cooling function strongly depends on the composition of the gas. The primordial composition of the gas is dictated by big bang nucleosynthesis. We take a primordial composition for the gas with 75 percent hydrogen and 25 percent helium by mass. In metal-free gas the dominant cooling process is atomic line cooling. As primordial gas is metal-free, we are implementing a zero metallicity cooling function by \cite{1993ApJS...88..253S}. At higher temperatures of $\rm 10^{6}-10^{8}$ K cooling is due to free-free transitions. For temperatures between $\rm 10^{4}-10^{6}$ K cooling is mainly due to HI and HeII Lyman alpha.

\begin{figure}
\centering
\includegraphics[scale=0.45]{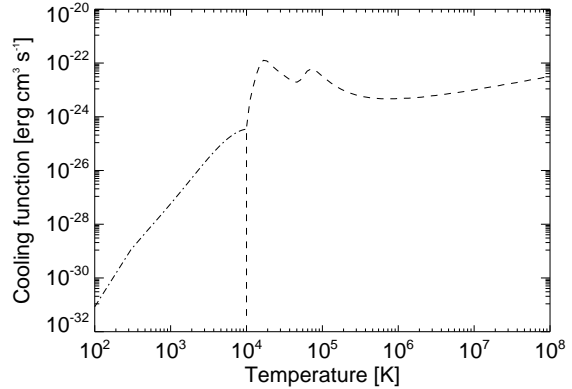}

\caption{Cooling function for zero metallicity. The dashed line shows the zero metal cooling function while the dash-dot line depicts the contribution due to molecular hydrogen.}
\label{fig:figure 1}
\end{figure}

Evolution of primordial density fluctuations is controlled by the ability of gas to cool down to low temperatures. The temperature of gas remains $\rm \approx$ 8000 K in the absence of $\rm H_{2}$ and HD \citep{1998A&A...335..403G} due to high excitation energies of primordial gas atoms. In the absence of metals, $\rm H_{2}$ cooling is crucial to form stars in $\rm T_{vir} > 10^{4}$ K halos. Vibrational and rotational modes of $\rm H_{2}$ can be excited at low temperatures $\rm \sim 500-1000 K$. Trace amounts of molecular hydrogen can form via gas phase reactions in the post-recombination era \citep{1968ApJ...154..891P}. We use the \cite{1998A&A...335..403G} cooling function to include cooling due to molecular hydrogen. We take a universal abundance of $\rm H_{2}$, i.e., $\rm 10^{-3}$. The zero metal cooling function is shown in figure \ref{fig:figure 1}, details can be found in the above mentioned articles.

\subsection {Variation in the equation of state}

The equation of state plays a vital role in structure formation. Fragmentation of gas clouds is strongly dependent on the polytropic equation of state (EOS) \citep{2003ApJ...592..975L}. The relation of the polytropic index of the EOS ($\rm \gamma$) to logarithmic derivatives of heating and cooling function implicitly includes radiative transfer effects \citep{2000ApJ...538..115S}. Under conditions of thermal equilibrium, by balancing the heating and cooling terms, the polytropic exponent which measures the compressibility of gas can be written as \citep{2002MNRAS.332..769S}:
\begin{equation}
\gamma=1+ {dlog T \over dlog \rho}  .
\end{equation}
\cite{2006ApJ...652..902S} have studied the effect of trapping Lyman alpha photons on the polytropic EOS for primordial halos. They found that trapping Lyman alpha photons stiffens the polytropic exponent significantly above unity. Consequently, the Jeans mass is boosted and fragmentation is halted.

As the collapse proceeds gas becomes optically thick. The escape probability of Lyman alpha photons, which perform a random walk both in space and frequency, is reduced and ultimately photons are trapped inside the cloud. If the photon travel time becomes larger than the free-fall time, the radiative energy is re-deposited inside the gas and consequently cooling is suppressed and the temperature of the gas becomes $\rm \geq 10^{4}$ K. The formation of molecular hydrogen will not take place as it is collisionally dissociated at temperatures higher than 8000 K. \cite{1977MNRAS.179..541R} have studied the influence of trapping radiation on collapsing gas clouds and concluded that trapping of Lyman alpha photons can inhibit fragmentation. They also mentioned that trapped radiation would exert pressure which is half of the pressure required to support the cloud and will not halt the overall collapse. Radiation pressure produced by trapping of Lyman alpha photons can considerably constrain the efficiency of star formation \citep{2002ApJ...569..558O}.

At optical depths $\rm \tau_{0} > 10^{7}$ the photon escape time $\rm t_{ph}$ becomes larger than $\rm t_{ff}$. Due to the weak dependence of $\rm t_{ph}$ on the number density ($\rm t_{ph} \propto n^{-1/9}$), the photon escape time remains longer than the free fall time ($\rm t_{ph} \propto n^{-1/2}$) during the collapse \citep{2006ApJ...652..902S}. The number of photons that is trapped at line center thus increases and the emission of Lyman alpha photons from the core tends to zero. Consequently, trapping becomes more effective during the collapse. The scattering of Lyman alpha photons during the collapse can introduce extra pressure which slows down the process of collapse. This changes the density of trapped Lyman alpha photons and induces fluctuations in the EOS which are not well captured in our case. A temporary softening of the EOS would favor fragmentation and our approach therefore gives an upper limit on the effect of Lyman alpha trapping. The minimum hydrogen column density required for the trapping is $\rm N_{H}=10^{21} cm^{-2}$.

By trapping Lyman alpha photons the polytropic index $\rm \gamma $ increases from 1.01 to 1.50 for hydrogen number densities of 1 to $\rm 10^{5} cm^{-3}$ \citep{2006ApJ...652..902S}. We found that high columns in our case stiffen the EOS at lower density values than mentioned by \cite{2006ApJ...652..902S}. Due to high columns, gas becomes opaque at lower number densities and causes more efficient trapping of Lyman alpha photons. Therefore, we have modified \cite{2006ApJ...652..902S} criteria for the change in polytropic index to include high column opacity effects. Their model is based on analytical results and does not take into account the resolution constraints. We use the formula given below to implement the change in the polytropic exponent of the EOS $\rm \gamma$,
\begin{figure}
\centering
\includegraphics[scale=0.45]{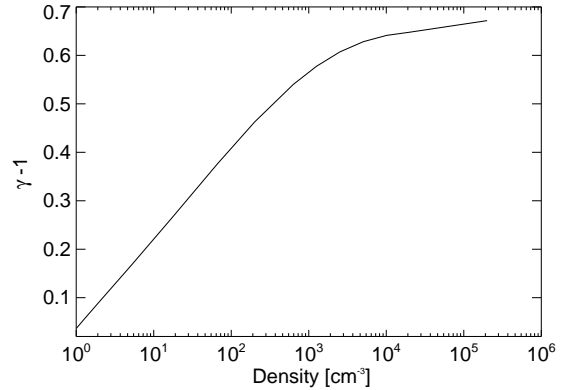}
\caption{Variation of gamma with density. The change in polytropic index ($\rm \gamma-1$) is plotted as a function of the number density ($\rm cm^{-3}$) of the gas. The figure shows the stiffening of the EOS by trapping Lyman alpha photons.}
\label{fig:figure 2}
\end{figure}
\begin{equation}
\gamma -1 \approx -{{1\over2} + {7\over 18}Bn_{1}^{{7/ 18}}\over {log(Cn_{1}^{0.5}) + Bn_{1}^{{7/ 18}}}},
\end{equation}
where the value of B $\rm \approx$ 0.47 $\rm cm^{7/6}$, C $\rm \approx$ $\rm 10^{-34} cm^{3/2}$ and $\rm n_{1}$ is 100 times the number density. Figure \ref{fig:figure 2} illustrates the stiffening of the EOS as a function of number density. It can be seen from the figure that gas remains close to isothermal up to 10$\rm~cm^{-3}$ and at higher densities opacity effects stiffen the value of $\rm \gamma$ well beyond unity. An increase in the value of $\rm \gamma$ enhances temperature well above $10^{4}$ K. Even a value of $\rm \gamma \sim 1$ should already be sufficient to halt fragmentation and will lead to isothermal collapse. Cooling due to helium becomes important at temperatures higher than 5.0 $\rm \times 10^{4}$ K. Above $\rm 10^{3} cm^{-3}$ opacity effects due to thermal bremsstrahlung trapping come into play and cooling continues to be suppressed. A value of $\rm \gamma \geq 4/3$ leads to an adiabatic collapse. Helium opacity effects can be included by making corrections to equation 2. Details can be found in \cite{2006ApJ...652..902S}.

\section {Results}

Our simulations show that density perturbations decouple from the Hubble flow and start to collapse through gravitational instability. The fate of these perturbations depends on the cooling and dynamical time scales. At low densities cooling is very inefficient. We see in our simulations that initially the gas is shock heated at low densities ($\rm \leq 10^{-1} cm^{-3}$). Virialization processes transform gravitational potential energy into kinetic energy of the gas and dark matter. As dark matter cannot dissipate, for gas this energy is converted into internal energy through shocks which raises the temperature of the gas to the virial temperature. We start with an adiabatic equation of state as perturbations in the inter-galactic medium are adiabatic in nature. According to linear perturbation theory, the EOS should show a power-law behavior in the low density regime \citep{1997MNRAS.292...27H,2002MNRAS.333..649S}, given by the equation

\begin{equation}
T=T_{0}{(\rho /\rho_{0})^{\gamma-1}}
\end{equation}
where $T_{0}$ and $\rho_{0}$ are the cosmic mean density and temperature. Our results agree with this analytical temperature density relation for the intergalactic medium.

At redshift 15, gas begins to cool and collapses in the gravitational potential provided by the dark matter. At densities of the order of $\rm 10^{4}$ times the mean density of the Universe cooling due to atomic lines (hydrogen and helium) becomes very efficient and gas becomes isothermal with a value of gamma of unity. Gas continues to collapse if  the cooling time is smaller than the dynamical time. We have investigated the dynamic and thermal evolution of primordial gas which is cooled or heated and collapses under different post-shock heating mechanisms. In the next subsections the results are discussed for each case individually. We use the isovolume connected components finding algorithm implemented in the visualization program Visit \citep{Childs:2005:ACS} to find fragments. For a given density threshold, we find the fragments and calculates their masses.

\subsection{Atomic line cooling}

\begin{figure*}
\centering
\begin{tabular}{c c}

\begin{minipage}{8cm}
\includegraphics[scale=0.32]{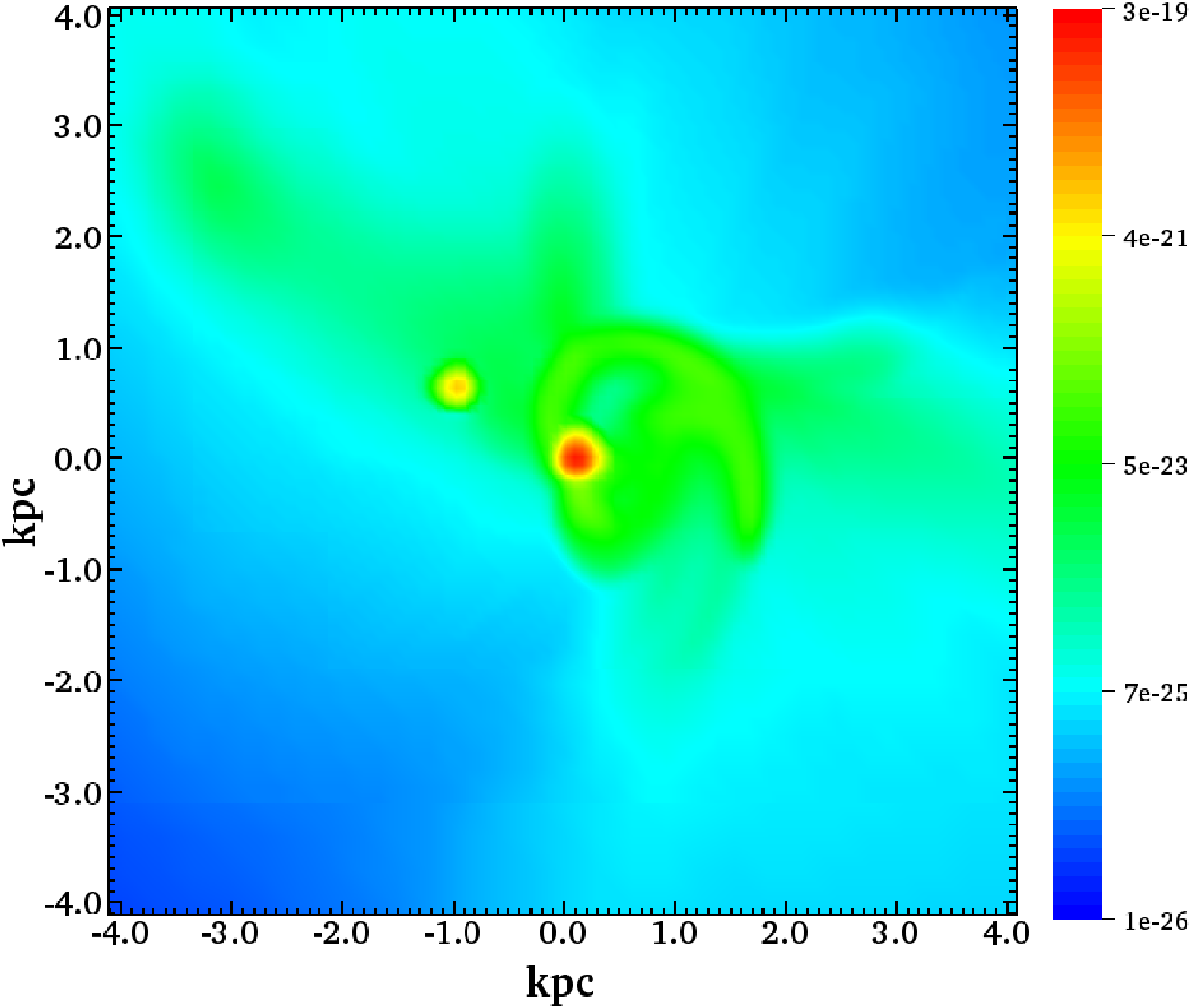}
\caption{Density slice through through the center of a halo for the atomic cooling case. Values corresponding to each color are shown in color bar, where blue color represents the lowest density and red represents the highest density. Distance scales are in comoving units. Here 1kpc in comoving units corresponds to 106 pc in proper units.}
\label{fig:denslice}
\end{minipage} &

\begin{minipage}{8cm}
\includegraphics[scale=0.32]{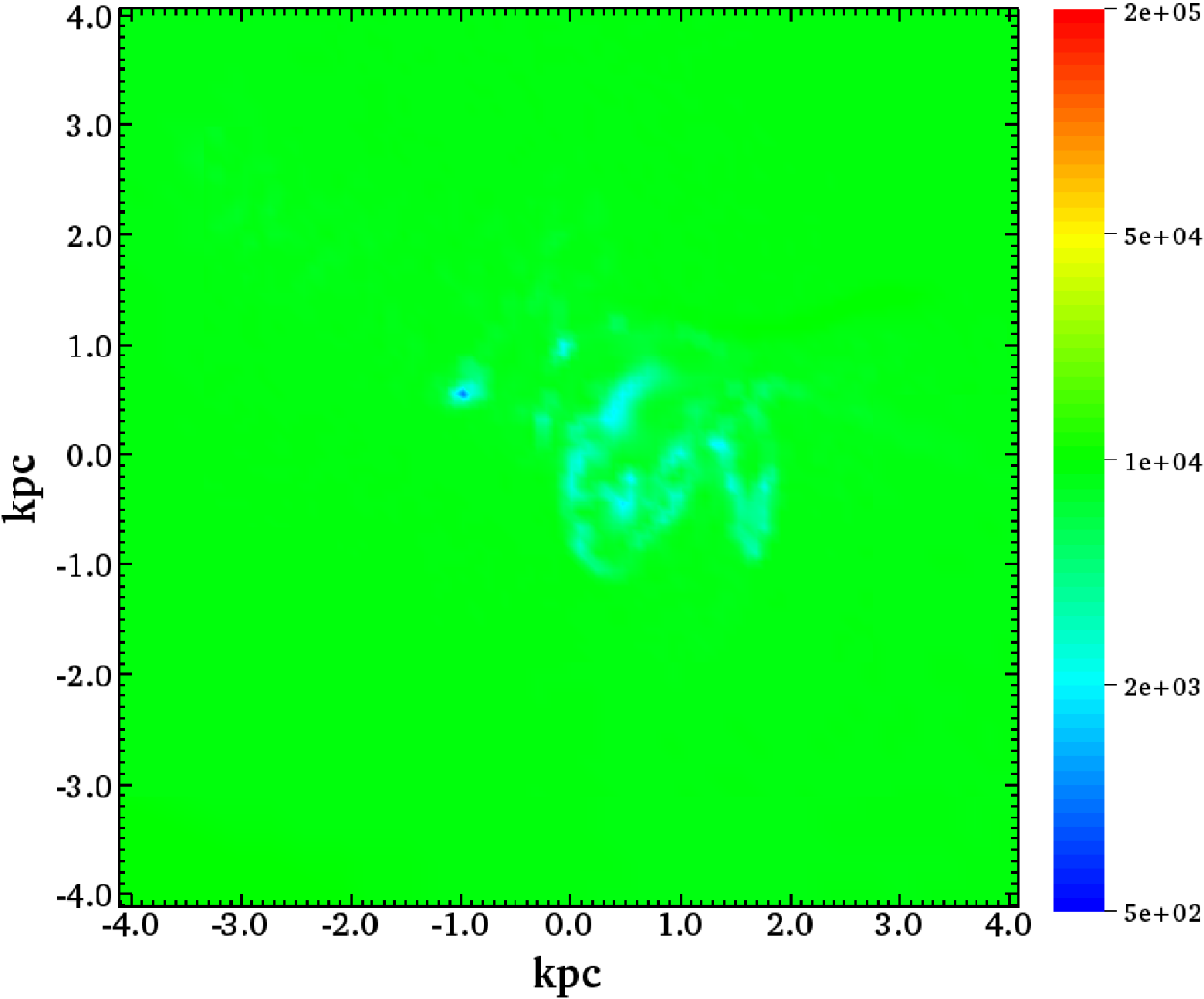}
\caption{Temperature slice through the center of a halo for the atomic cooling case. The figure shows the temperature corresponding to density slice of figure \ref{fig:denslice}. Blue color represents the lowest temperature and red represents highest temperature.}
\label{fig:atomictmp}
\end{minipage} \\

\begin{minipage}{8cm}
\includegraphics[scale=0.28]{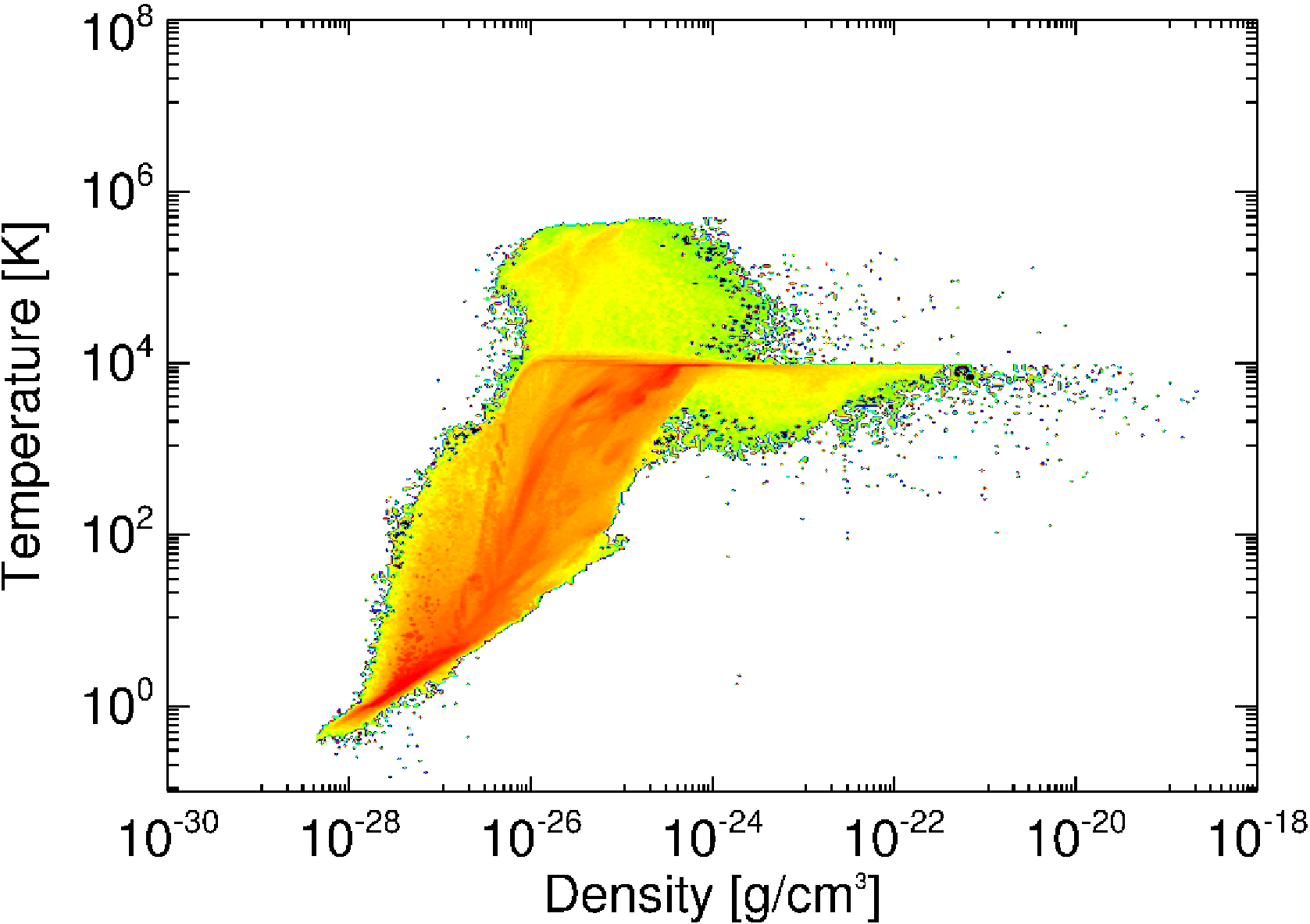}
\caption{Density-temperature phase diagram for the atomic cooling case. Temperature is plotted against volume weighted density in filled color contours by merging adaptive grid into uniform grid, where red color shows the highest mass and purple shows lowest mass.}
\label{fig:atomicphase}
\end{minipage}&

\begin{minipage}{8cm}
\includegraphics[scale=0.32]{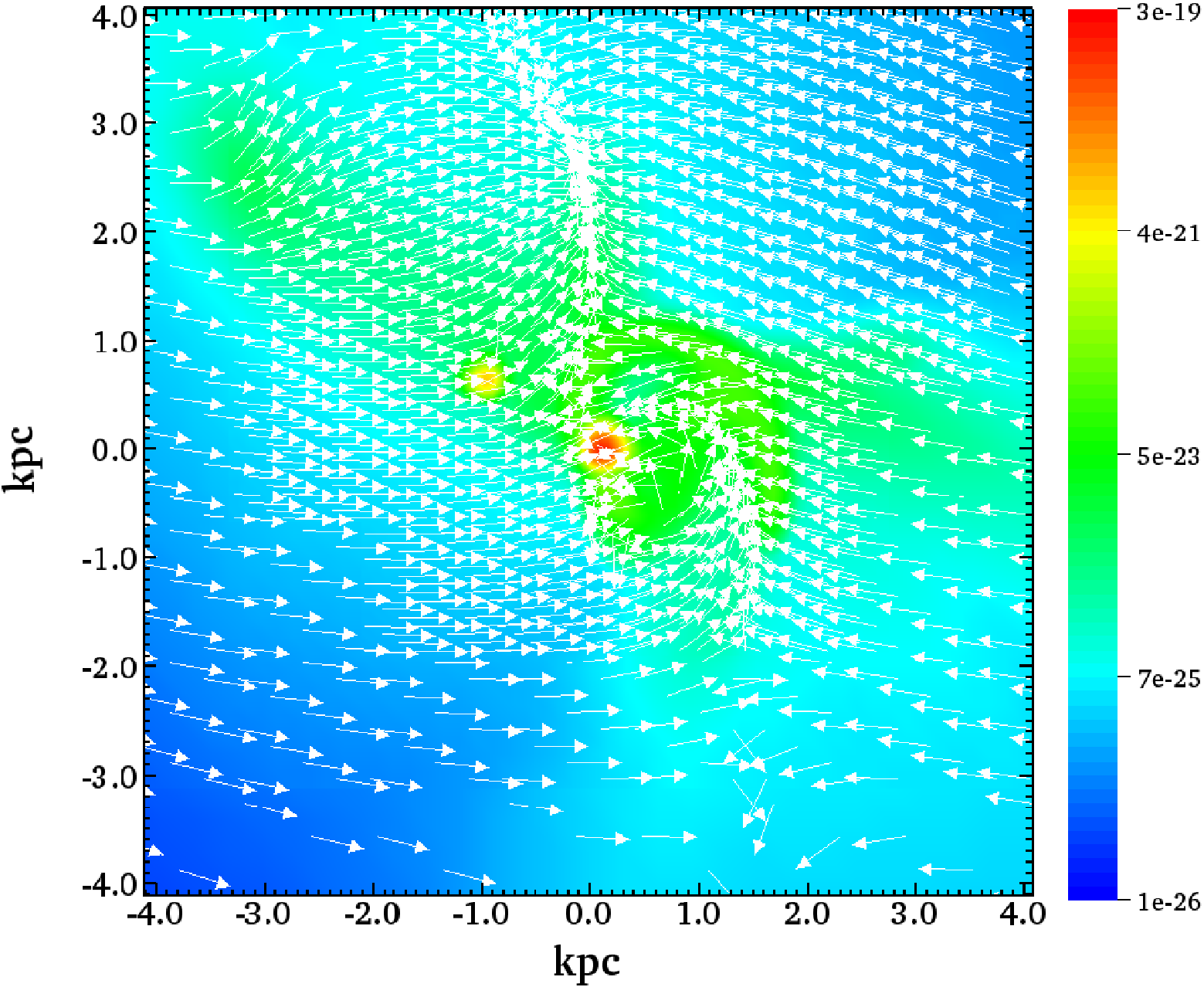}
\caption{ Velocity vectors are overplotted on density slice for the atomic cooling case. Velocity vectors are indicated by white color arrows. Head of the arrows indicates the direction of flow.}
\label{fig:atomicvel}
\end{minipage}

\end{tabular}
\end{figure*}
\begin{figure*}
\centering
\begin{tabular}{c c}

\begin{minipage}{8cm}
\includegraphics[scale=0.28]{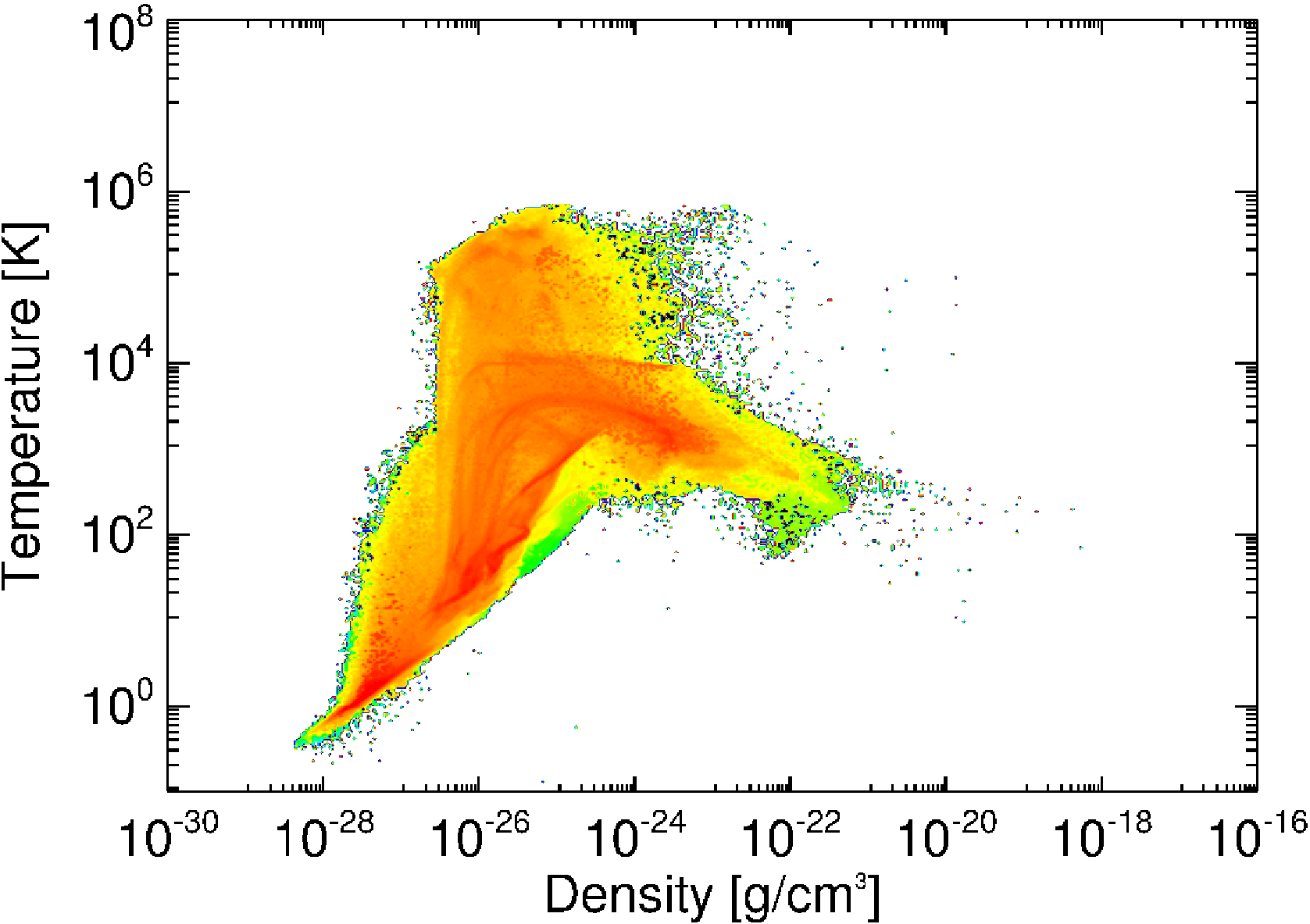}
\caption{Density-temperature phase diagram for the molecular cooling case. Temperature is plotted against the volume weighted density in filled color contours by merging adaptive grid into a uniform grid, where red color shows the highest mass and purple shows lowest mass.}
\label{fig:h2phase}
\end{minipage} &

\begin{minipage}{8cm}
\includegraphics[scale=0.32]{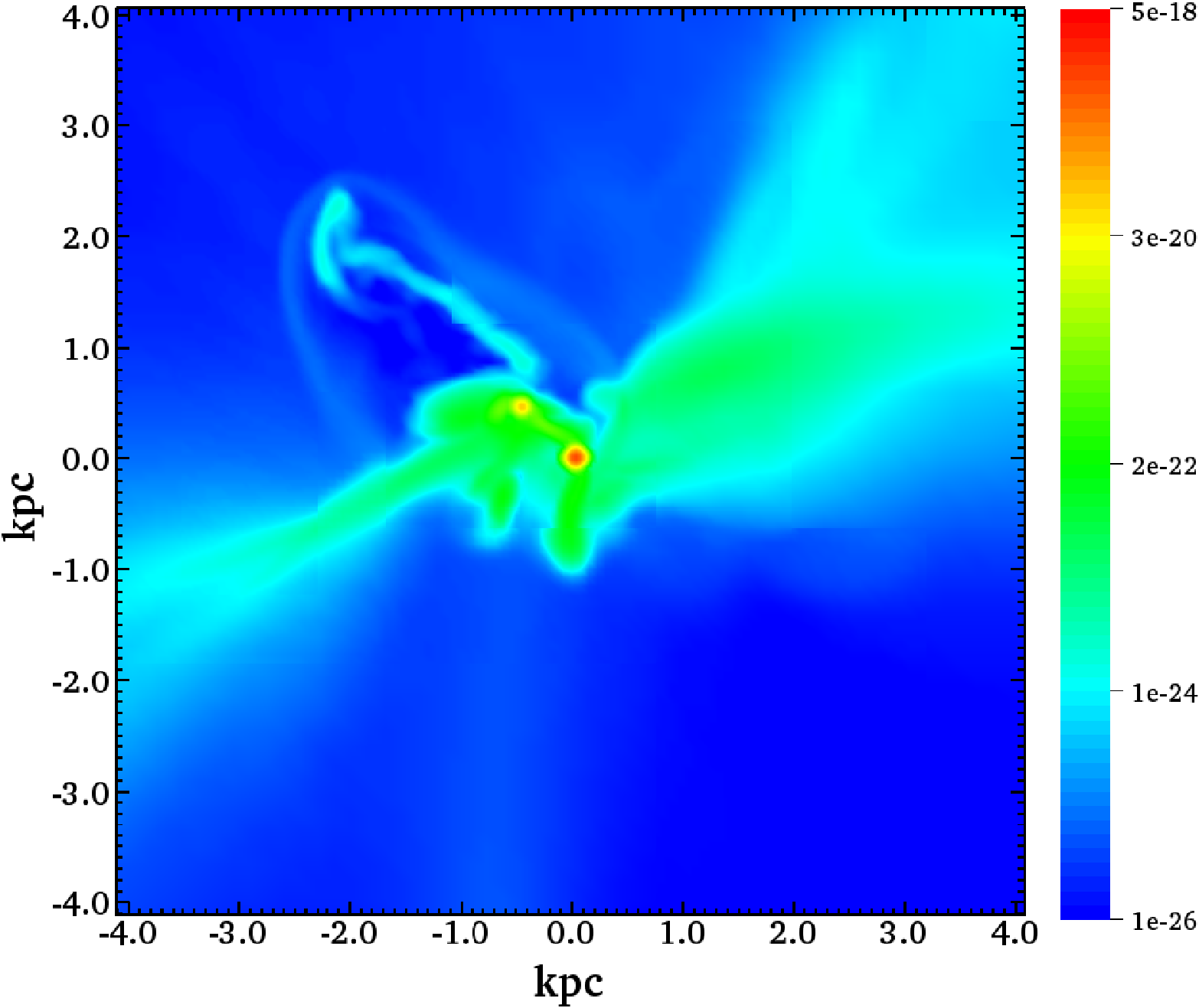}
\caption{Density slice through the center of a halo for the molecular cooling case. Values corresponding to each color are shown in the color bar, where blue color represents the lowest density and red represents the highest density. Distance scales are in comoving units. Here 1kpc in comoving units corresponds to 106 pc in proper units.}
\label{fig:h2slice}
\end{minipage} \\

\begin{minipage}{8cm}
 \includegraphics[scale=0.3]{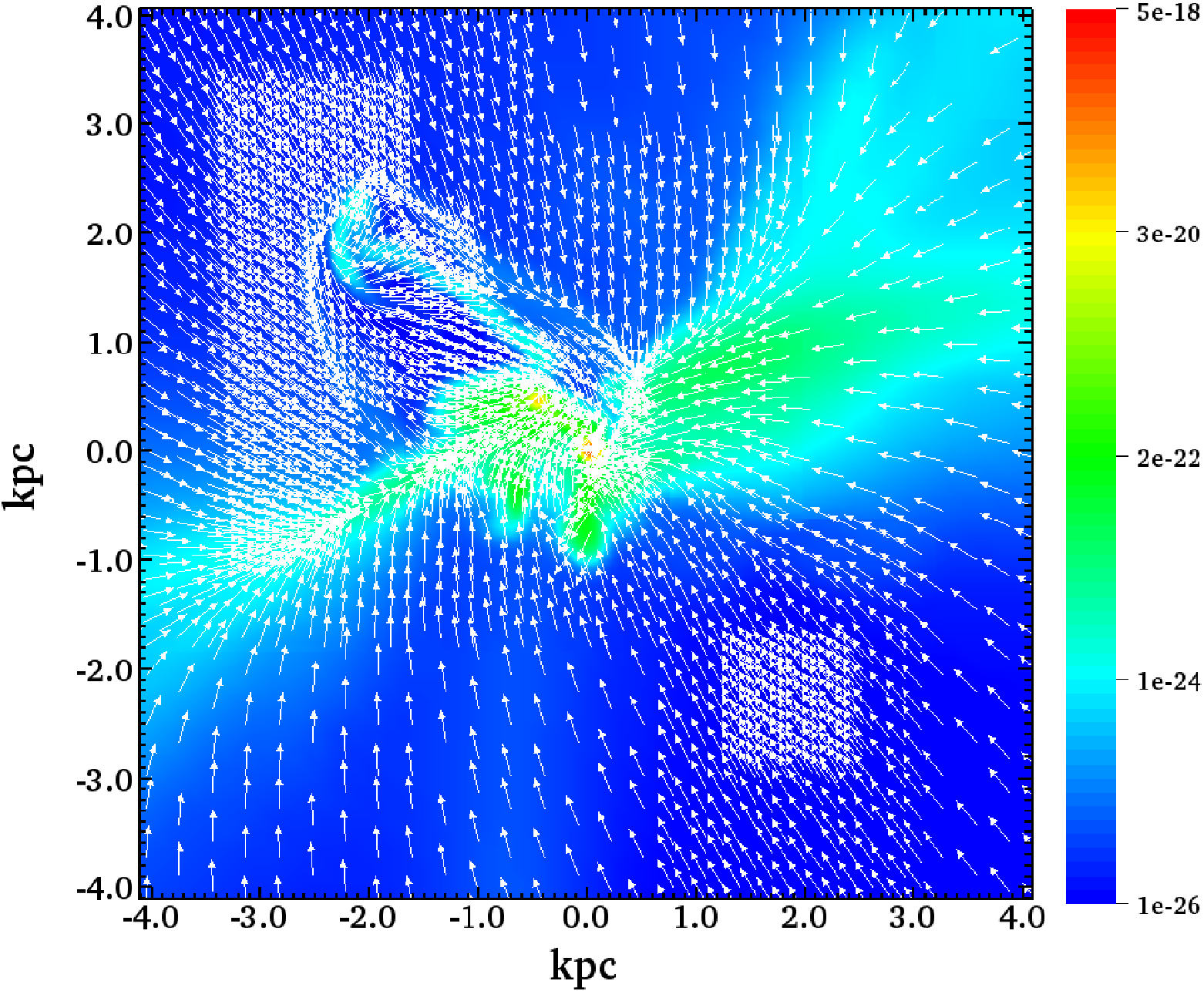}
\caption{ Velocity vectors are overplotted on the density slice for the molecular hydrogen cooling case. Velocity vectors are indicated by white color arrows. The head of the arrows indicates the direction of flow.}
\label{fig:h2vel}
\end{minipage}&

\begin{minipage}{8cm}
\includegraphics[scale=0.32]{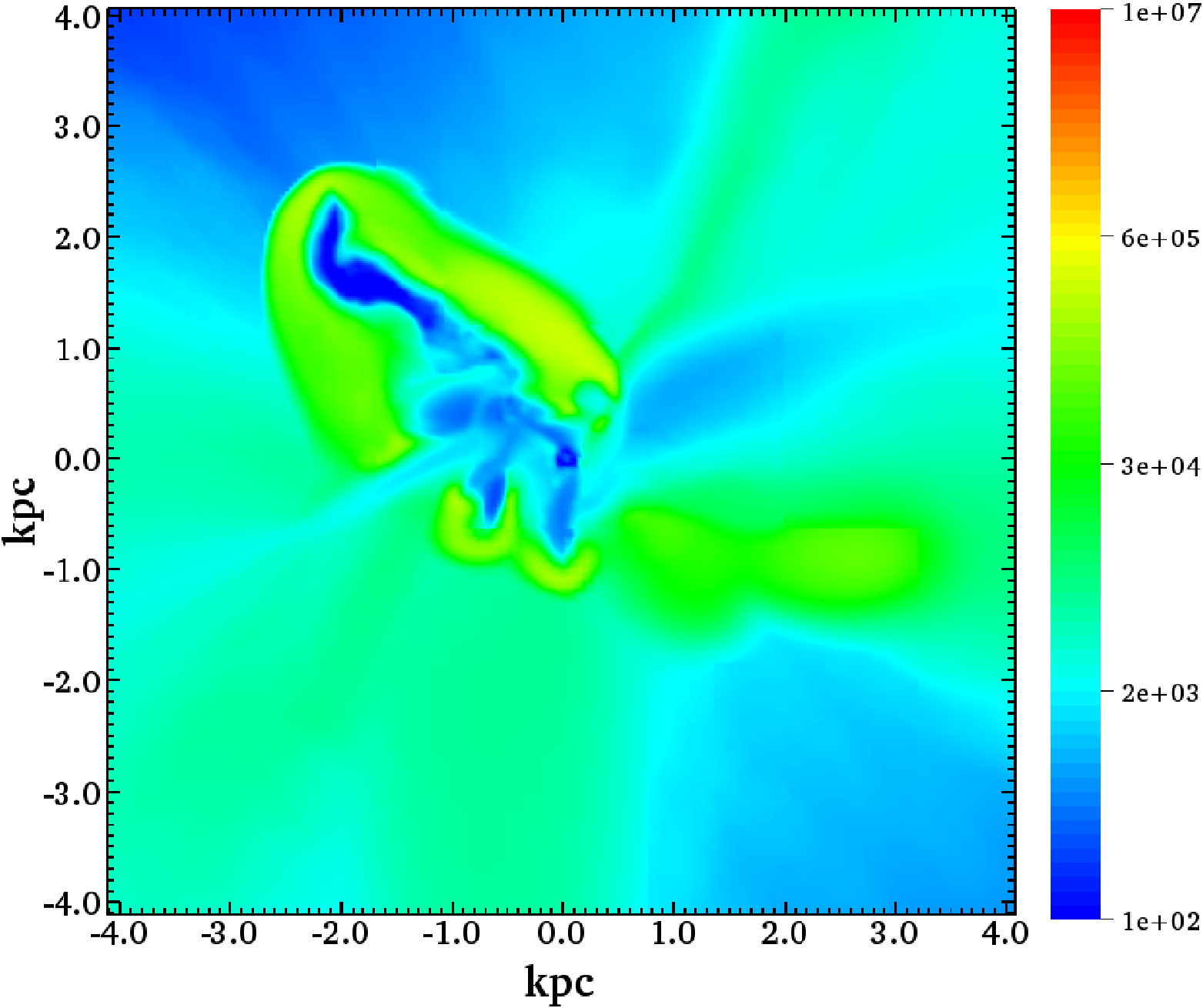}
\caption{Temperature slice through the center of a halo for the molecular hydrogen cooling case. The figure shows the temperature corresponding to the density slice of figure \ref{fig:h2slice}. The temperature of fragments is close to 100 K which can be seen in blue color in the plot.}
 \label{fig:h2tmp}
\end{minipage}
\end{tabular}
\end{figure*}

Metal-free halos with virial temperatures $\rm >10^{4}$ K and no Lyman alpha trapping can be prone to molecular hydrogen formation if not irradiated by an intense UV flux. There may exist some halo pairs where molecular hydrogen is photodissociated by Lyman Werner radiation of the other halo \citep{2008MNRAS.391.1961D}. Recent calculations by \cite{2010MNRAS.402.1249S} indicate that specific intensities required to suppress the formation of molecular hydrogen are 3-10 times lower than previous estimates of $\rm J_{crit} \sim$ 1000 in units of $\rm 10^{-21}$ erg $\rm cm ^{-2} s^{-1}Hz^{-1} sr ^{-1}$. Consequently, the population of such halos is enhanced by a factor of $\rm \approx 10^{3}$. We assume that the halo is metal free and illuminated by intense UV flux. Hence, the formation of $\rm H_{2}$ does not take place. Initially gas heats up to its virial temperature. When the cooling time becomes less than dynamical time, it begins to cool and collapse in dark matter potentials. As cooling due to Lyman alpha photons is very efficient, the gas cannot virialize by gaining internal energy, so its kinetic energy must be increased to reach virial equilibrium. Consequently, during virialization the gas becomes turbulent. It continues to cool and collapse into the center of a halo.

Gas in the halo collapses up to densities of a few times $\rm 10^{5} cm^{-3}$. It fragments into two massive clumps of $\rm 2.6\times 10^{6} M_{\odot} ~and~ 1.3\times 10^{6} M_{\odot}$ as shown in figure \ref{fig:denslice}. Collapse is nearly isothermal and gas in the halo is cooled down to only a small fraction of the virial temperature. Isothermal behavior is depicted in temperature slice in figure \ref{fig:atomictmp}. The temperature remains around  $\rm \sim 8000$ K. The value of the polytropic index remains close to one. Isothermal collapse is preventing it to fragment into smaller clumps and the Jeans mass is $\rm 2\times10^{5}M_{\odot}$.

The gas passes through different phases during collapse which are shown in the temperature-density phase diagram in figure \ref{fig:atomicphase}. It is initially shock heated at low densities and subsequently cooled by atomic line cooling. Most of its mass is lying at low densities and only a small fraction ($\rm <1$\%) is cooled and collapsed into clumps. It falls into the center of a halo through filaments and the direction of flow is shown by velocity vectors in figure \ref{fig:atomicvel}. The typical velocity is around 100 $km~s^{-1}$. We also examined the entropy of gas in the halo and found that it is $\rm 10^{-2} keV cm^{-2}$.

Our results are in agreement with previous results \citep{2003ApJ...596...34B,2008ApJ...682..745W}. We do not see disk formation in our case as mentioned by \cite{2009MNRAS.393..858R}. The reason for this disagreement could be due to difference in resolution. We do not exclude the possibility of disk formation which may form after collapse. Inefficient fragmentation of a halo for isothermal collapse is according to the expectation of theoretical models. 

\subsection{Molecular hydrogen cooling}

\begin{figure*}
\centering
\begin{tabular}{c c}

\begin{minipage}{8cm}
\includegraphics[scale=0.28]{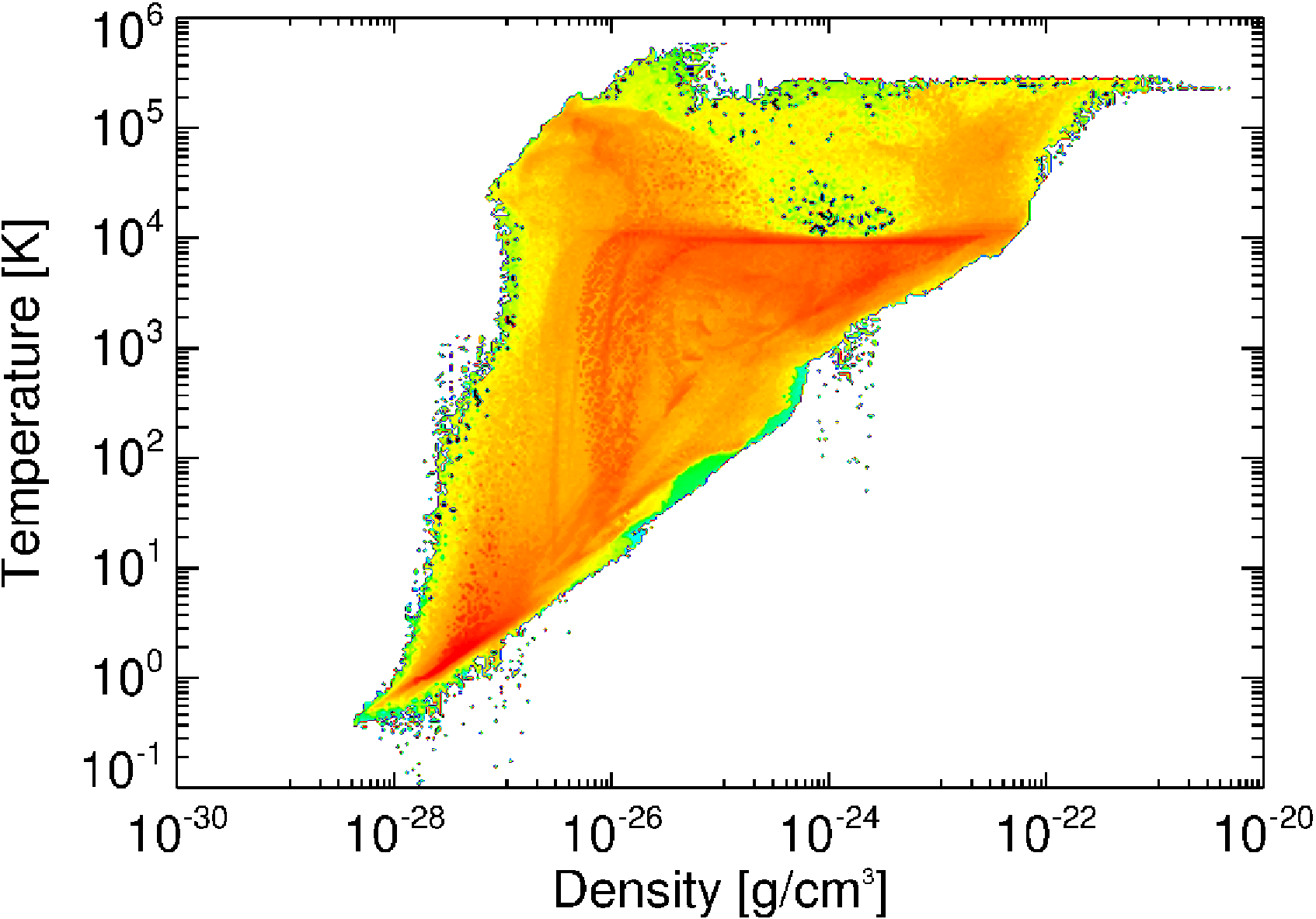}
\caption{Density-temperature phase diagram for the Lyman alpha trapping case. Temperature is plotted against volume weighted density in filled color contours by merging adaptive grid into uniform grid, where red color shows the highest mass and purple shows the lowest mass.}
\label{fig:lymanphase}
\end{minipage} &

\begin{minipage}{8cm}
\includegraphics[scale=0.32]{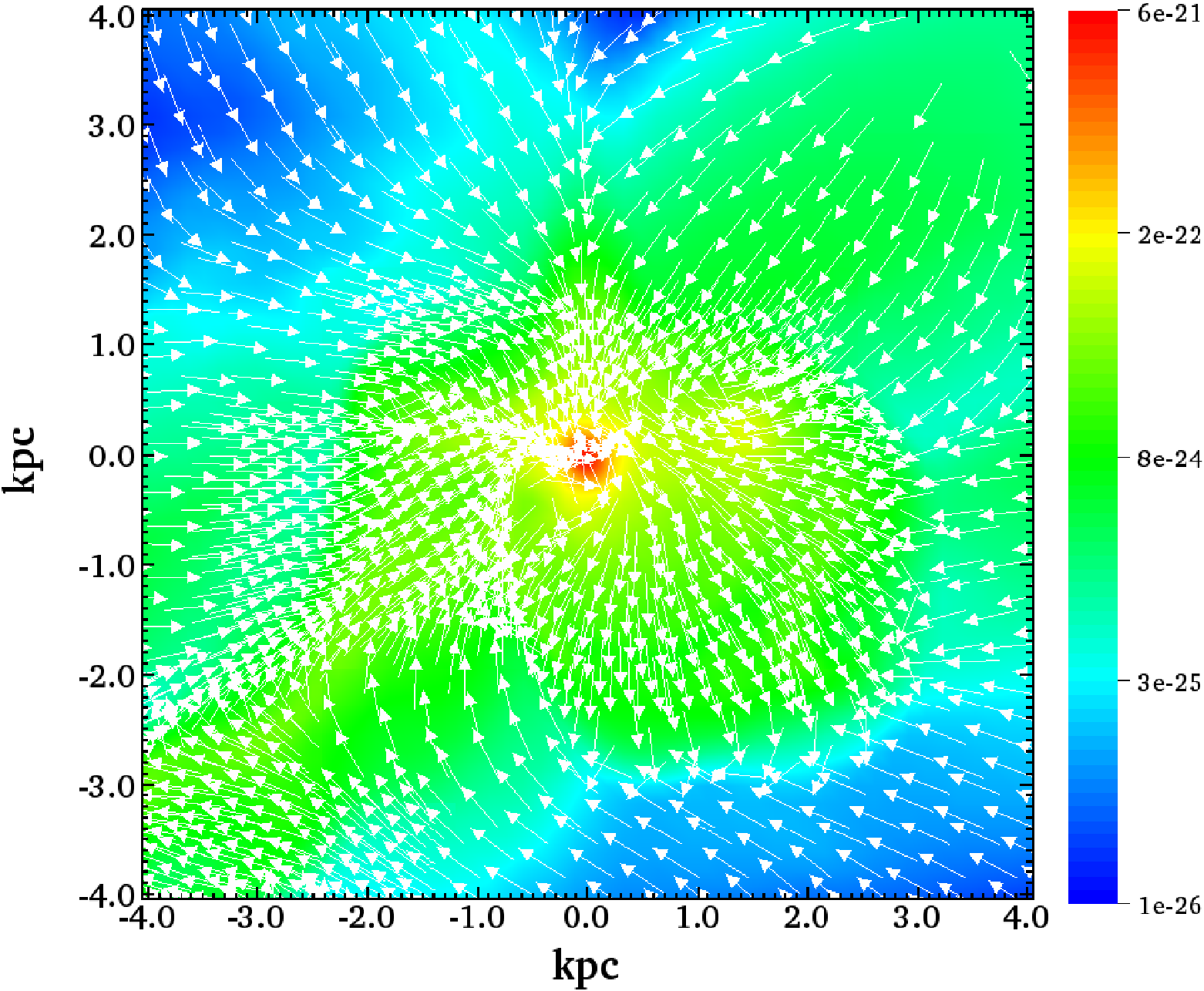}
\caption{ Velocity vectors are overplotted on density slice for the Lyman alpha trapping case. Velocity vectors are indicated by white color arrows. Head of the arrows indicates the direction of flow.}
\label{fig:lymanvel}
\end{minipage} \\

\begin{minipage}{8cm}
\includegraphics[scale=0.32]{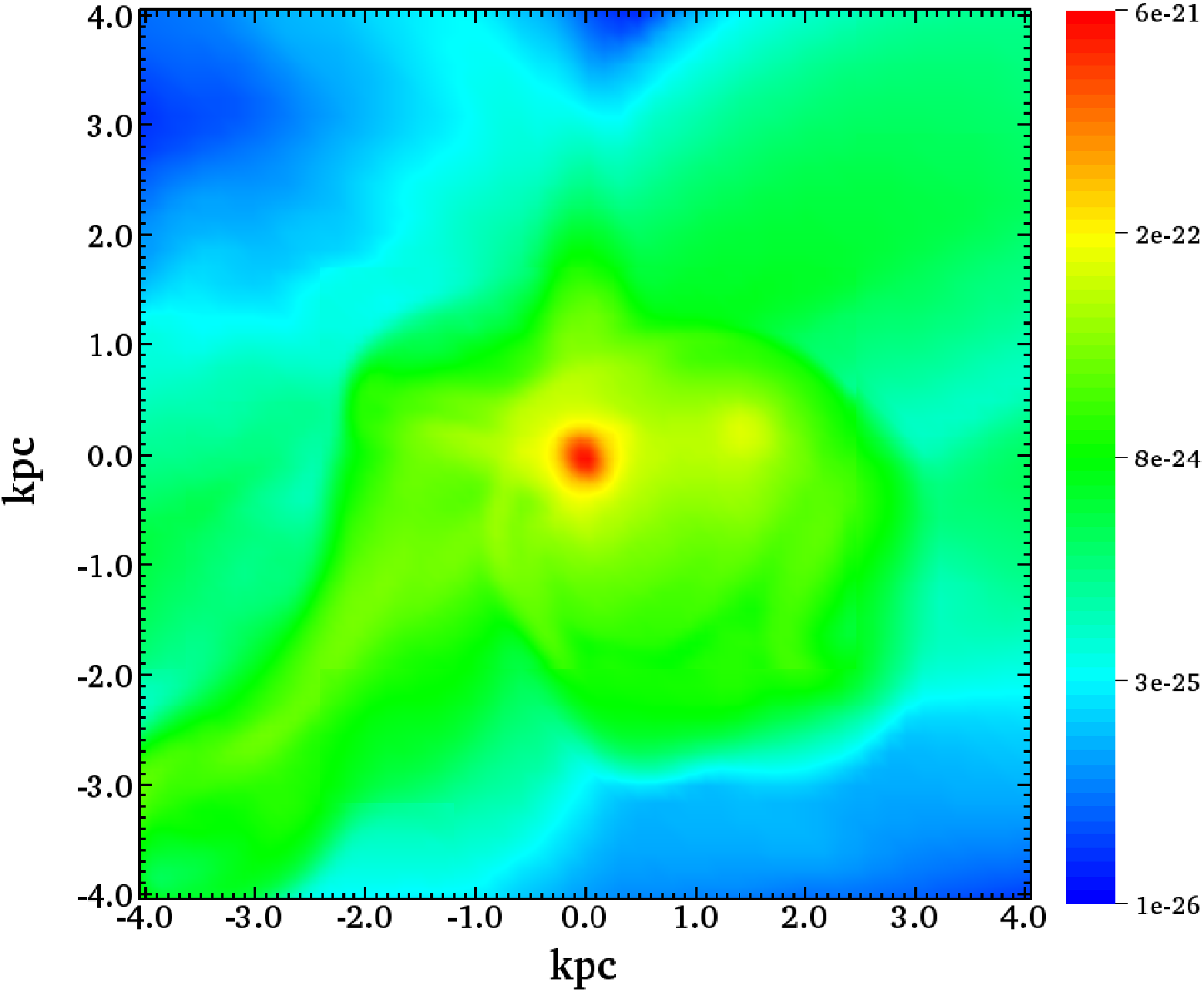}
\caption{Density slice through the center of a halo for the Lyman alpha trapping case. Density is shown in rainbow colors where red represents highest density and purple represents lowest density. Distance scales are in comoving units. Here 1kpc in comoving units corresponds to 106 pc in proper units.}
\label{fig:lymandens}
\end{minipage} &

\begin{minipage}{8cm}
\includegraphics[scale=0.32]{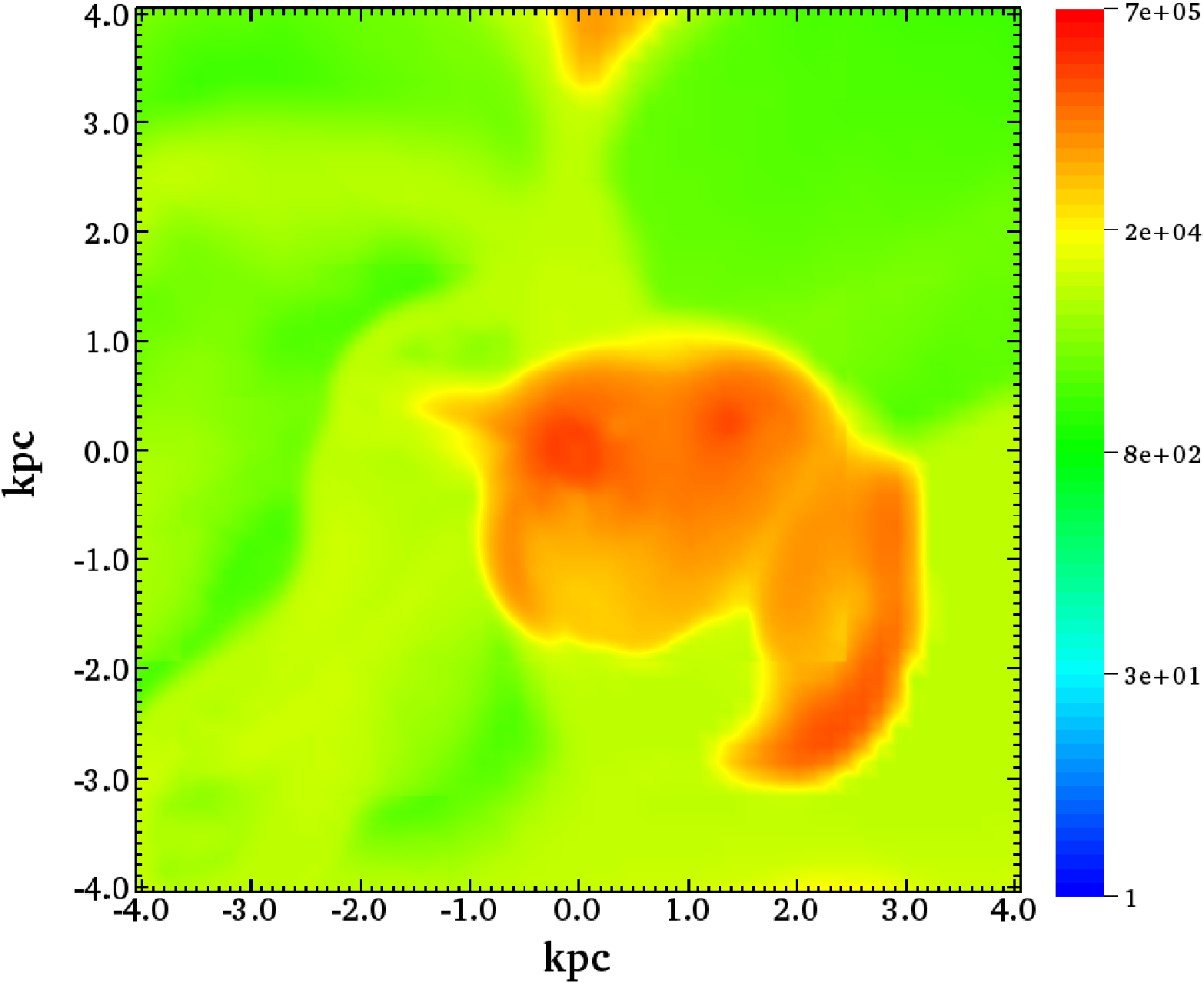}
\caption{Temperature slice through the center of a halo for the Lyman alpha trapping case. The figure shows the temperature corresponding to density slice of figure \ref{fig:lymandens}. Temperature of object is high compared to atomic cooling case. Distance scales are in comoving units.}
\label{fig:lymantmp}
\end{minipage}

\end{tabular}

\end{figure*}
Adding molecular hydrogen cooling enhances the low temperature cooling and increases the fragmentation. We see that gas is heated up to its virial temperature by shock waves at lower densities and subsequently cools by atomic line cooling. At densities higher than 1 $\rm cm^{-3}$ cooling due to molecular hydrogen comes into play and lowers the temperature down to $\rm \sim$ 100 K. During collapse it passes through different phases, its multiphase behavior is shown in the temperature-density phase diagram in figure \ref{fig:h2phase}. Most of the gas is lying at low densities and only a small fraction ($\rm <1$\%) is cooled and collapses into minihalos.

The equation of state is softened due to $\rm H_{2}$ cooling and the value of the polytropic index is lowered to less than unity. Hence, the Jeans mass is reduced. A density slice is shown in figure \ref{fig:h2slice}. $\rm H_{2}$ cooled gas is seen at the intersection of filaments. The gas collapses into 6 minihalos in contrast to the Lyman alpha trapping case, where gas does not collapse into minihalos. Their masses are $\rm 2\times10^{6}M_{\odot},8\times10^{5}M_{\odot},8\times10^{5}M_{\odot},3\times10^{5}M_{\odot},1\times10^{5}M_{\odot}~and~3.4\times10^{5}M_{\odot}$. Only two of the minihalos are visible in the slice. Their sizes are significantly smaller than the other cooling cases. The Jeans mass is reduced to $\rm \sim 10^{3} M_{\odot}$ and the gas collapses to densities of a few times $\rm 10^{6} cm^{-3}$. The gas becomes turbulent during virialization and falls into the center of a halo through cold streams as seen in \citep{2009MNRAS.395..160K,2001ApJ...562..605F,2009Natur.457..451D,2008MNRAS.387.1021G,2007ApJ...665..899W}. The density is $\rm \sim 10~cm^{-3}$ and temperature is $\rm \sim 300K$ in these streams. The direction of flow is presented in figure \ref{fig:h2vel}. The temperature of the minihalos is around 100 K and is shown in figure \ref{fig:h2tmp}. Shock heated gas is seen in the vicinity of the minihalos and the penetration of cold flows is even more prominent in this plot. The introduction of $\rm H_{2}$ cooling has significantly reduced the entropy of the gas down to $\rm 10^{-5}~keV cm^{-2}$. Entropy is low in the minihalos, higher in outskirts, and even boosted in shocks ($\rm 10^2~keV cm^{-2}$).

\subsection{Lyman alpha trapping}

It is generally assumed that gas, even though very optically thick ($\rm \tau > 10^7$) to scattering of Lyman alpha photons, can cool efficiently through Lyman alpha because radiation leaks out in the line wings. High columns of neutral atomic gas make the gas so optically thick that photon travel time exceeds the $\rm t_{ff}$. So, using the effectively optically thin cooling rate is not a good approximation above  $\rm 10^{20} cm^{-2}$. Consequently, the EOS stiffens as explained in the previous section and cooling is suppressed. Gas remains almost optically thin up to densities on the order of $\rm 10~cm^{-3}$. At densities $\rm \geq 10^{2}~cm^{-3}$ it becomes opaque enough that cooling due to Lyman alpha photons is quenched. When the gas density exceeds $\rm 10^{3}~cm^{-3}$, at typical columns of the order of $\rm 10^{23}~cm^{-2}$, the 2-photon continuum channel is shut down and photons are absorbed by neutral hydrogen in the surroundings \citep{2006ApJ...652..902S}. Under these conditions the halo collapses adiabatically to form a massive object.


\begin{figure*}
\centering
\begin{tabular}{c c}

\begin{minipage}{8cm}
\includegraphics[scale=0.32]{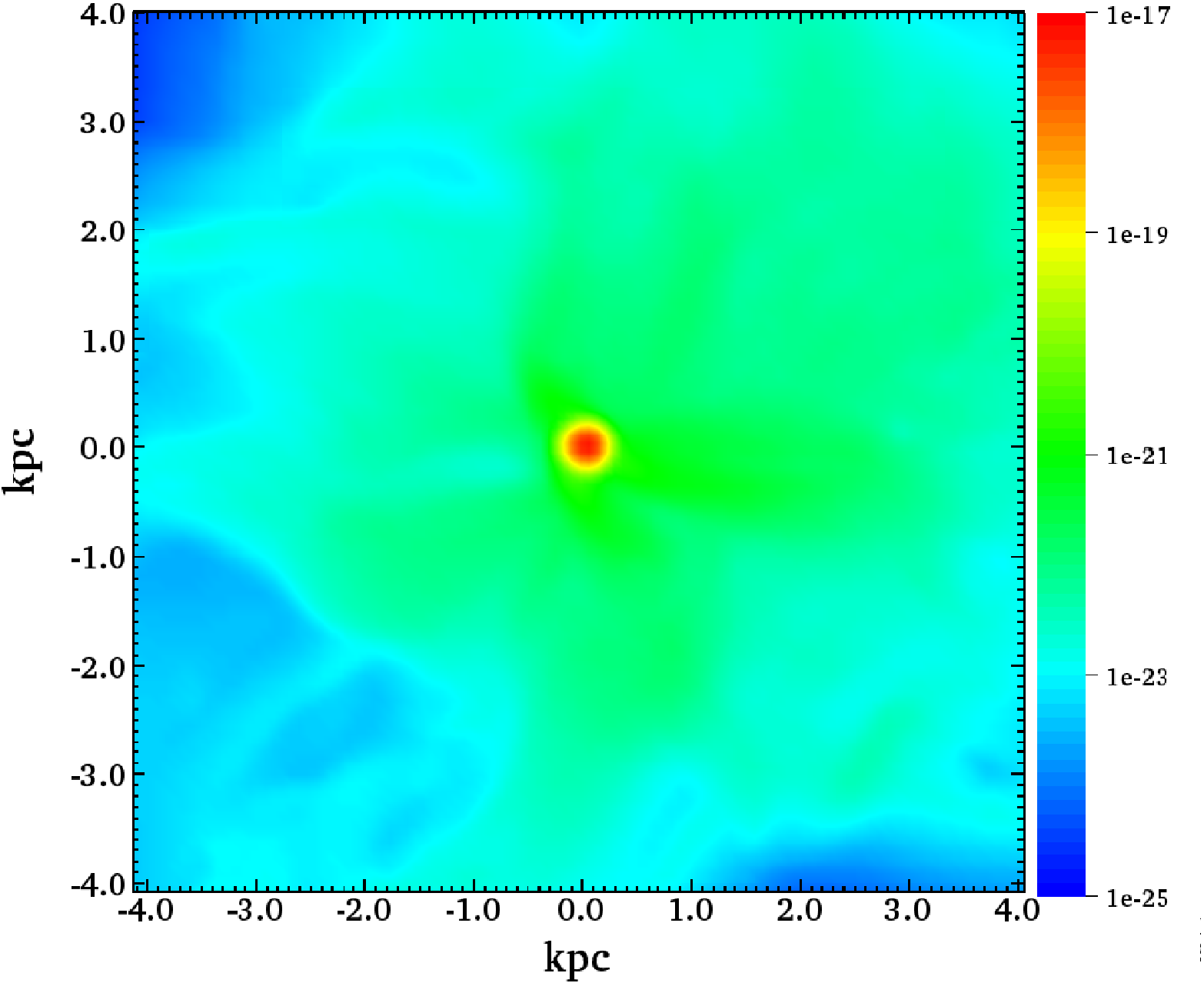}
\end{minipage} &

\begin{minipage}{8cm}
\includegraphics[scale=0.32]{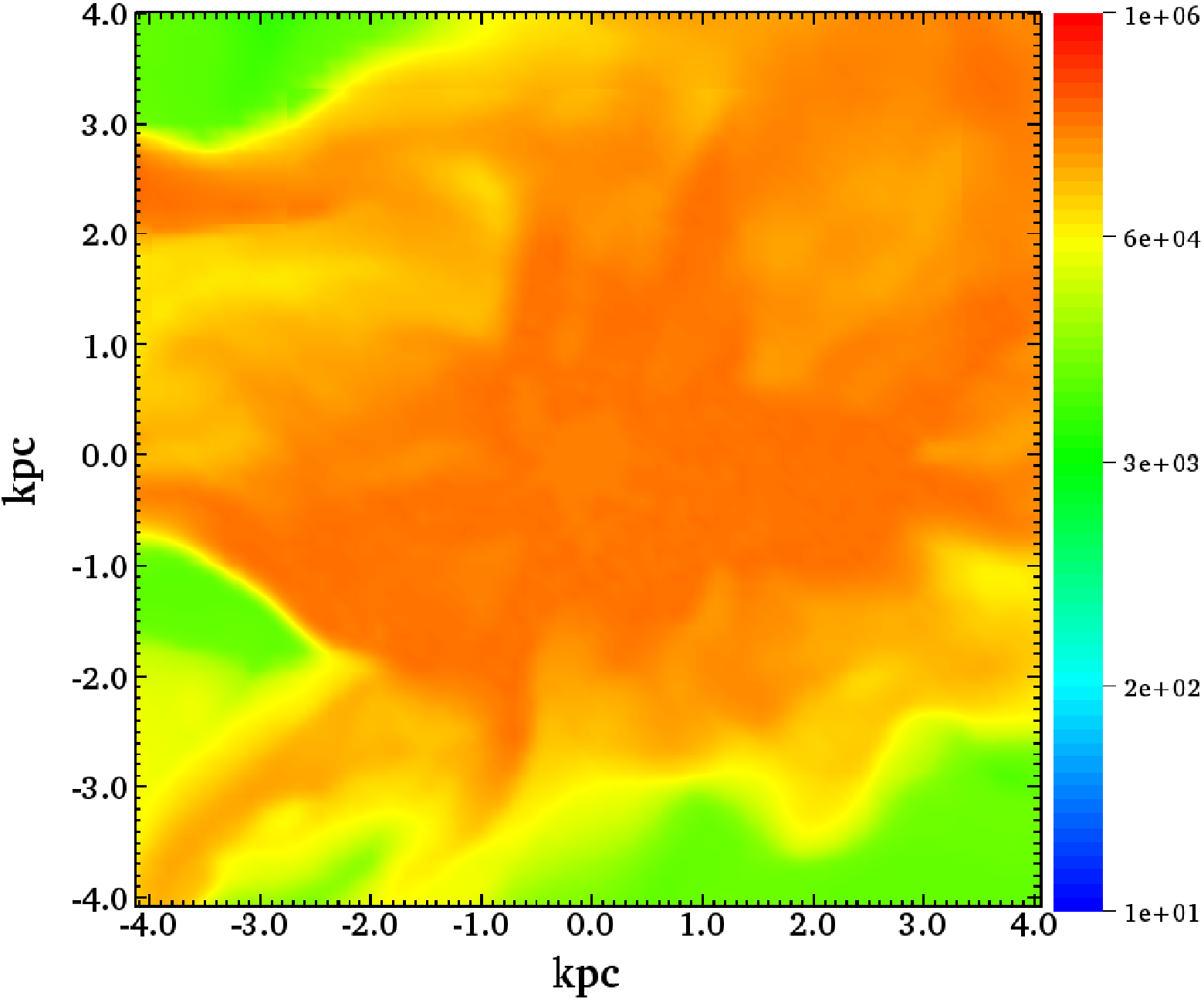}
\end{minipage} \\

\begin{minipage}{8cm}
 \includegraphics[scale=0.28]{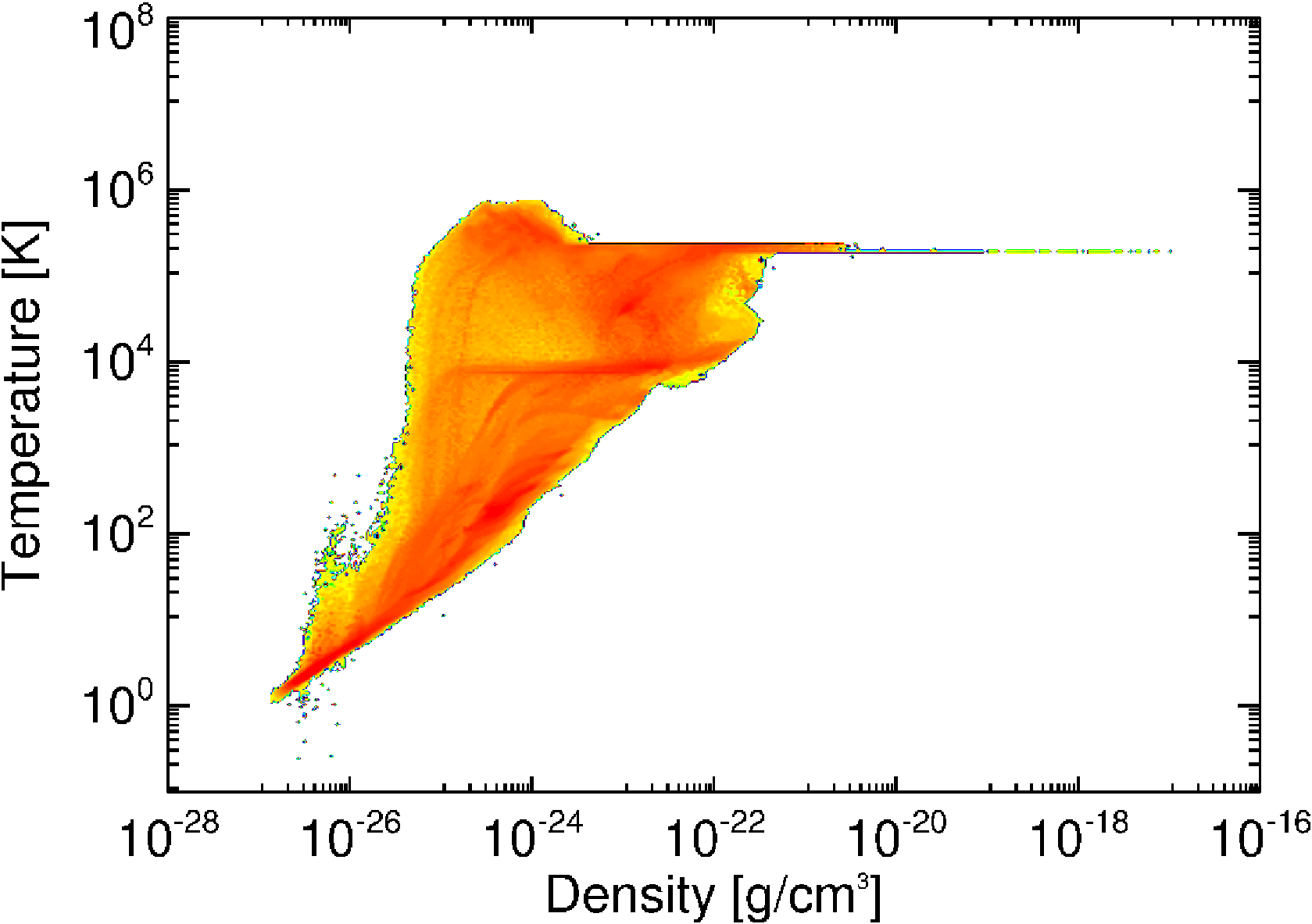}
\end{minipage}&

\begin{minipage}{8cm}
\includegraphics[scale=0.32]{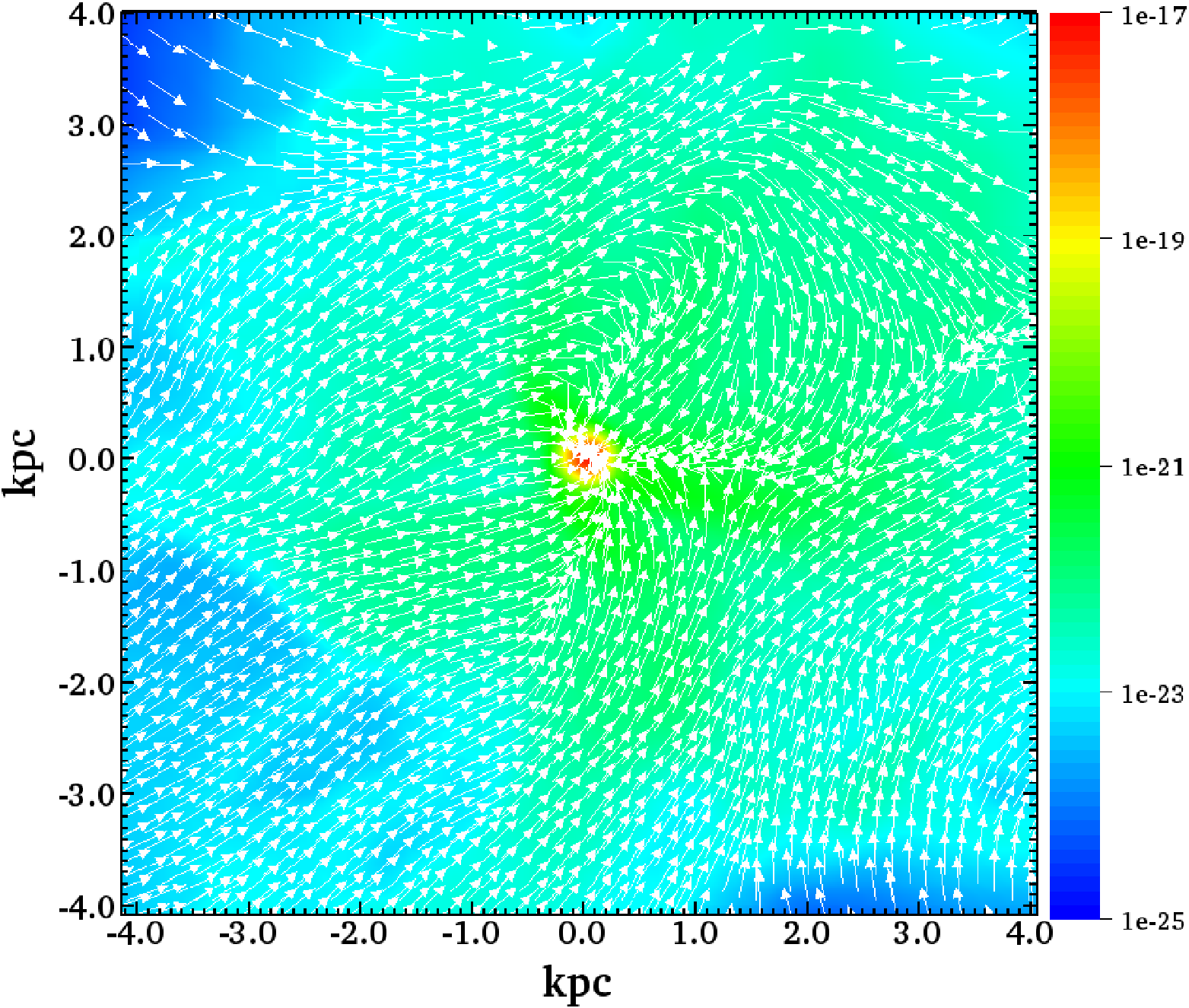}
\end{minipage}

\end{tabular}
\caption{ The figure represents a higher density peak halo for the Lyman alpha trapping case. The upper left panel shows the density slice through the center of a halo. The temperature slice corresponding to density slice is depicted in the upper right panel. The bottom left panel shows the phase diagram. The flow of gas is seen in the bottom right panel. Values corresponding to colors are shown in the color bars. Distance scales are in comoving units. Here 1kpc in comoving units corresponds to 53 pc in proper units.}
\label{fig:lymannew}
\end{figure*}

Gas is initially shock heated at low densities, becomes isothermal at densities $\rm \sim 1 cm^{-3}$, while at densities $\rm \geq 1 cm^{-3}$ EOS is stiffened because of Lyman alpha trapping. This multi-phase behavior is depicted in the density-temperature phase diagram in figure \ref{fig:lymanphase}. Most of the gas is lying at low densities and only a small fraction ($\rm <1$\%) collapses to high densities. Gas in the halo becomes turbulent during virialization and is segregated according to its angular momentum distribution (low angular momentum gas falls into the center). Hence, the collapse happens in the center of a halo. An object of mass $\rm 4.0\times10^{6}M_{\odot}$ is formed at the intersection of filaments which may lead to a massive object i.e. an IMBH. We see signs of a cold flow, which has already been seen in numerical simulations by \citep{2009Natur.457..451D,2009MNRAS.395..160K,2001ApJ...562..605F}. Baryons accumulate into the center of a halo via a two-phase medium by entrance of cold flow streams through the shock heated medium. These flows also explain the observed Lyman alpha blob morphologies \citep{2009arXiv0902.2999D}. The direction of flow is shown by velocity vectors in figure \ref{fig:lymanvel}. The values of the velocity in the center of halo are about 100 $\rm km ~s^{-1}$ which are typical values for such objects.

\begin{figure*}
\centering
\begin{tabular}{c c}

\begin{minipage}{8cm}
\includegraphics[scale=0.32]{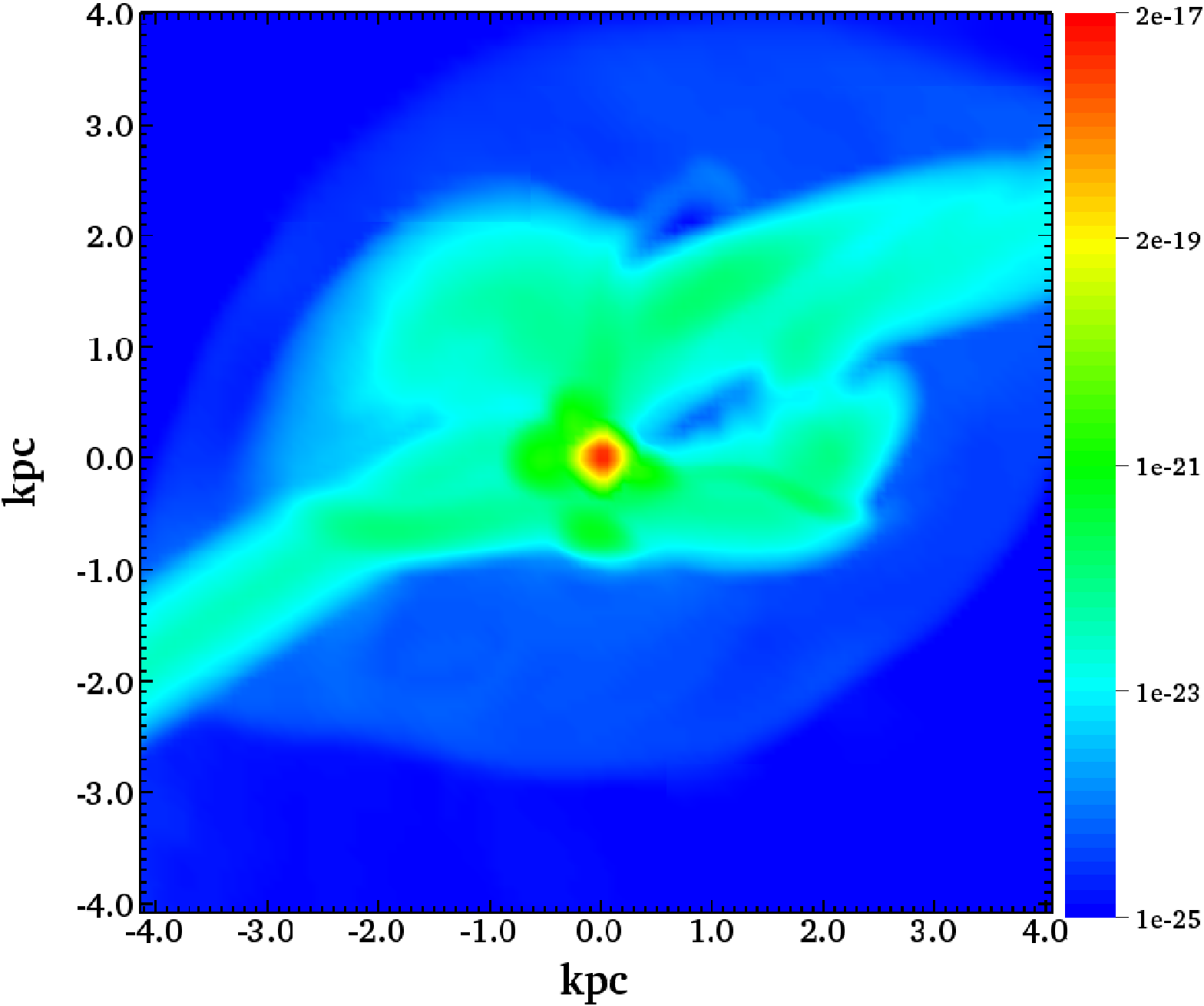}
\end{minipage} &

\begin{minipage}{8cm}
\includegraphics[scale=0.32]{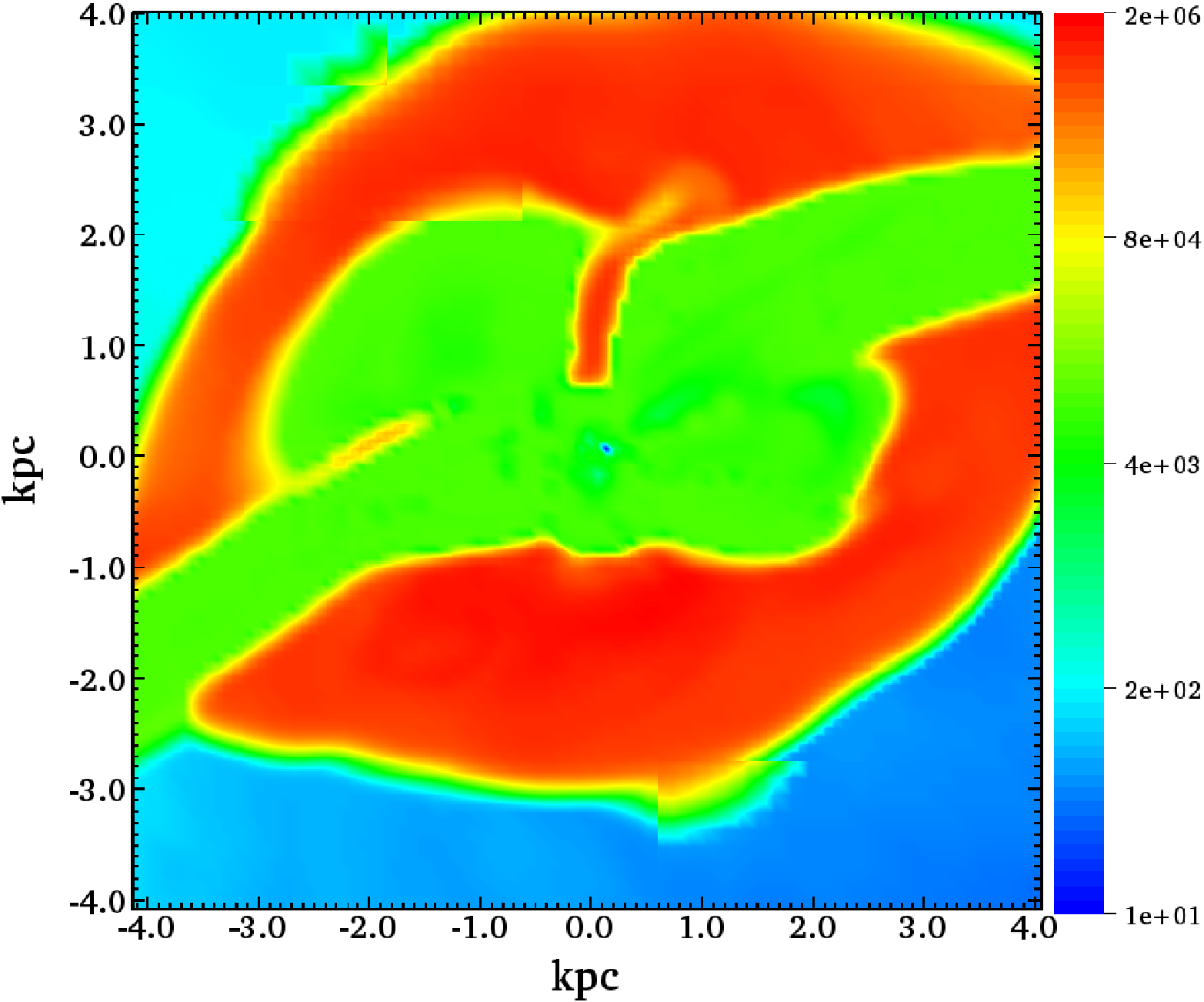}
\end{minipage} \\

\begin{minipage}{8cm}
\includegraphics[scale=0.28]{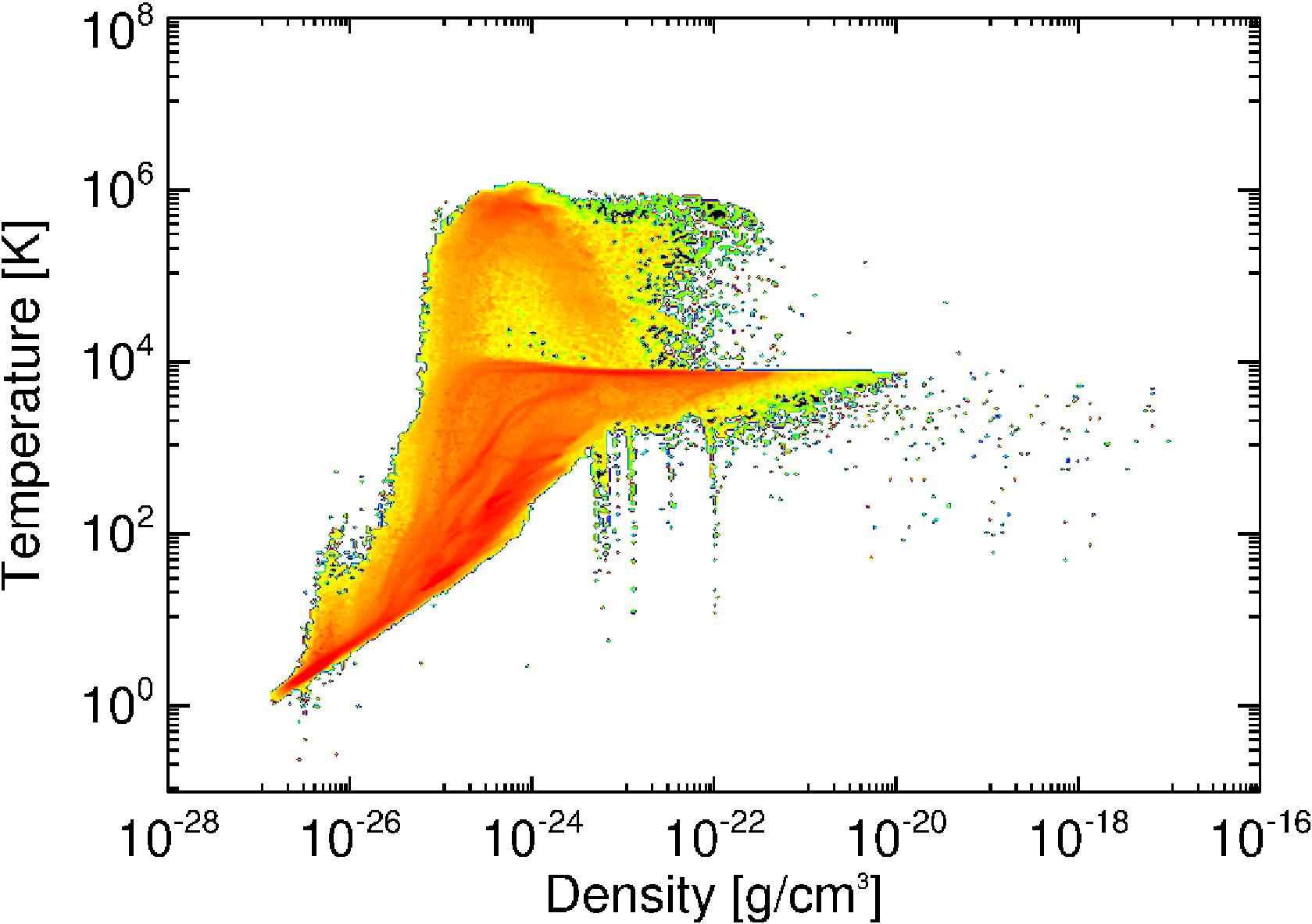}
\end{minipage}&

\begin{minipage}{8cm}
\includegraphics[scale=0.32]{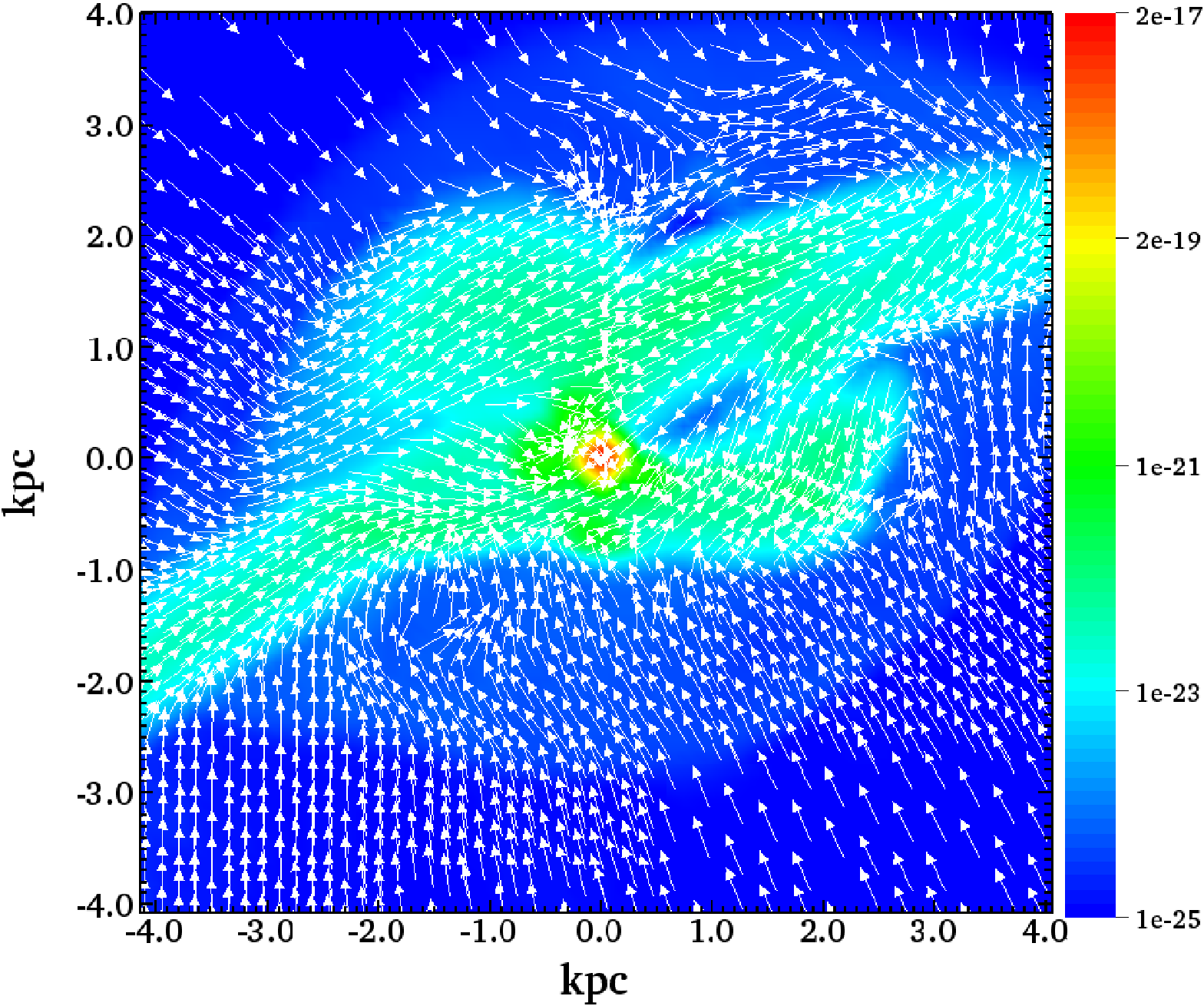}
\end{minipage}

\end{tabular}
\caption{ The figure represents a higher density peak halo for the atomic cooling case. The upper left panel shows the density slice through the center of a halo. The temperature slice corresponding to the density slice is depicted in the upper right panel. The bottom left panel shows the phase diagram. The flow of gas is seen in bottom right panel. Values corresponding to colors are shown in the color bars. Distance scales are in comoving units. Here 1kpc in comoving units corresponds to 53 pc in proper units.}
\label{fig:atomicnew}
\end{figure*}

\begin{figure*}
\centering
\begin{tabular}{c c}

\begin{minipage}{8cm}
\includegraphics[scale=0.32]{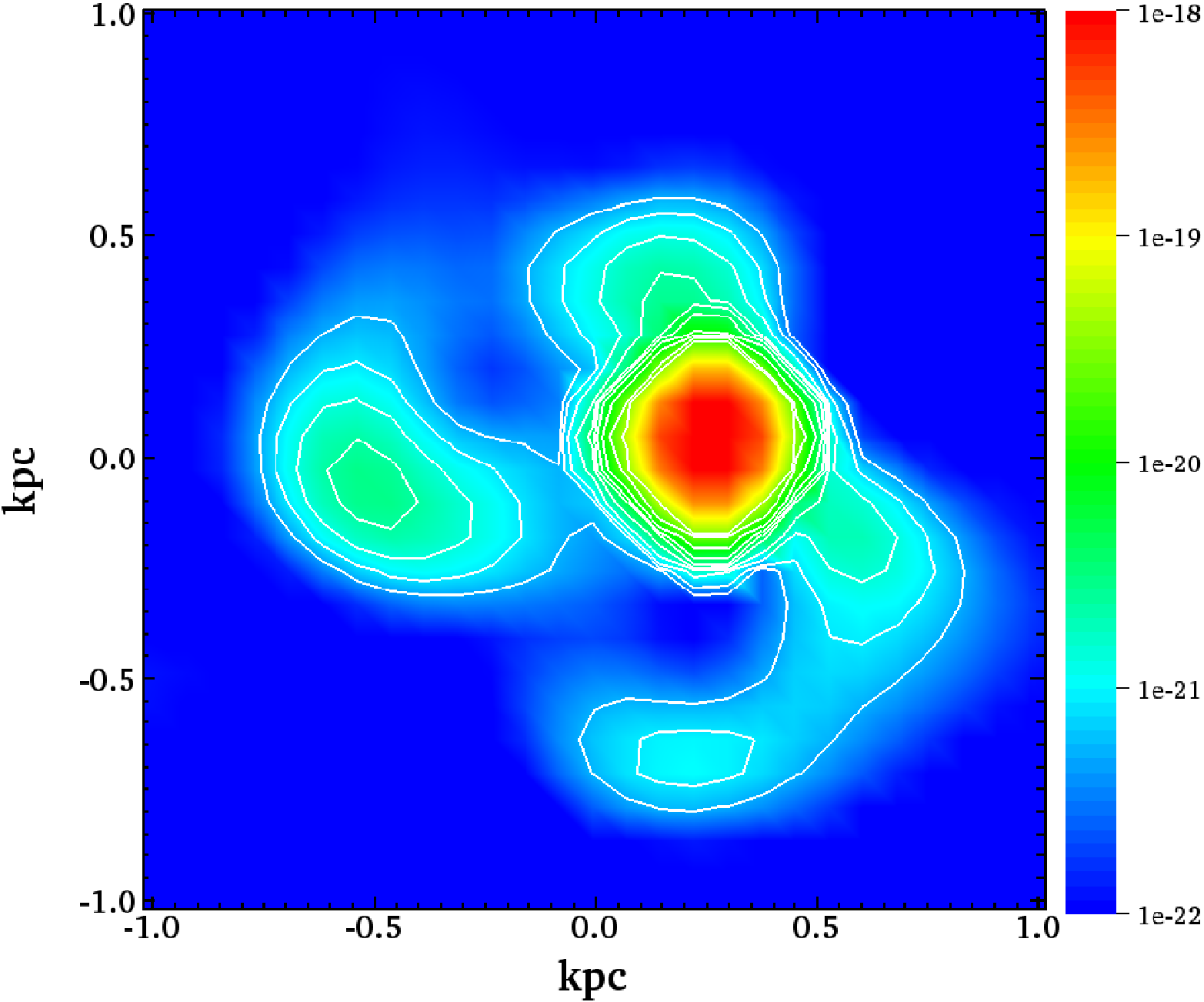}
\end{minipage} &

\begin{minipage}{8cm}
\includegraphics[scale=0.32]{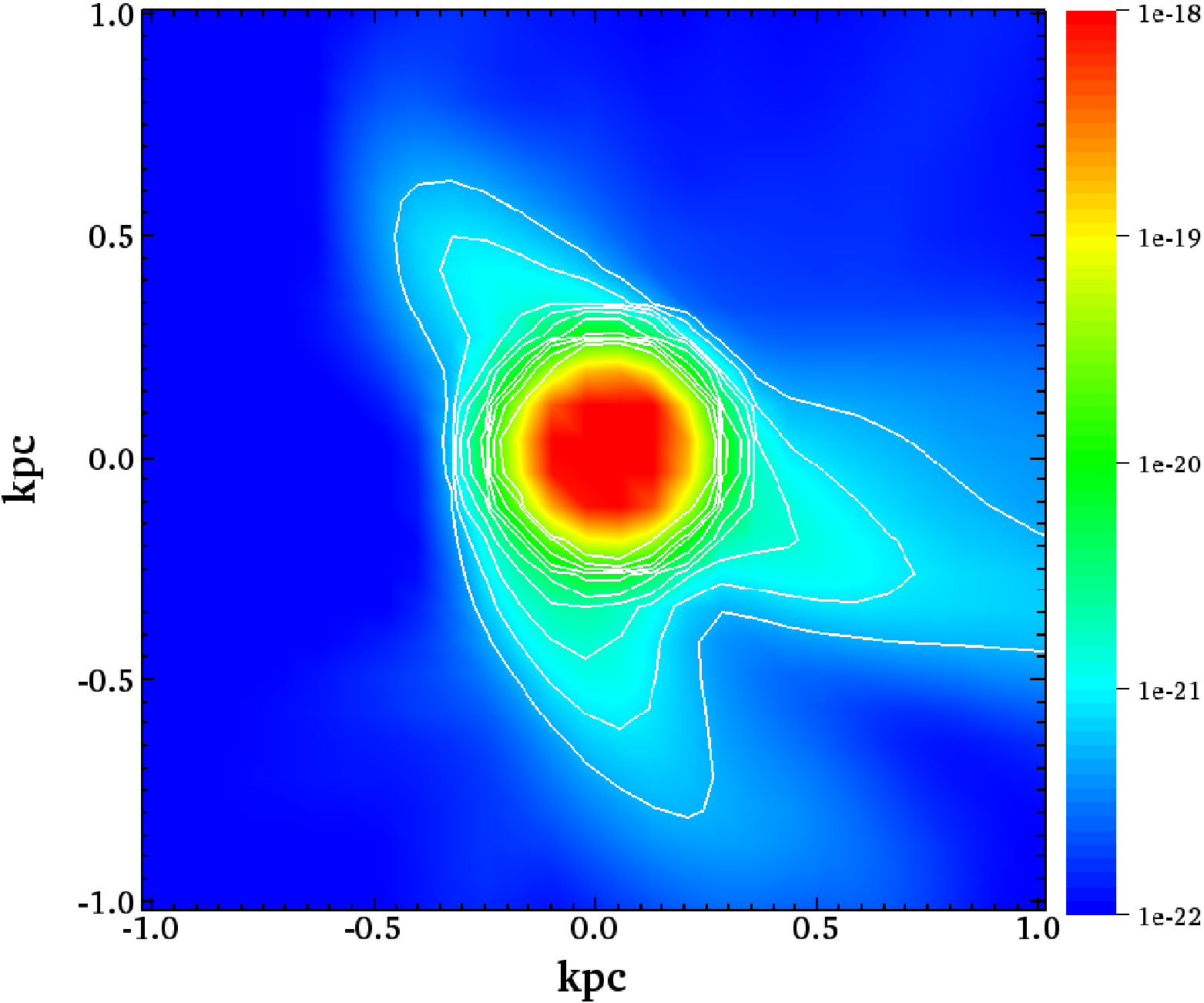}
\end{minipage} \\

\end{tabular}
\caption{Density slices of the inner 2 kpc region (zoomed) of the figures \ref{fig:atomicnew} (upper left panel) and figure \ref{fig:lymannew} (upper left panel). The left panel shows the density slice for the atomic cooling case. The right panel shows the density slice for the Lyman alpha case. Contours are overplotted to show the density structure. Distance scales are in comoving units. Here 1kpc in comoving units corresponds to 53 pc in proper units.}
\label{fig:new3}
\end{figure*}

\begin{figure*}
\centering
\begin{tabular}{c c}

\begin{minipage}{8cm}
\includegraphics[scale=0.32]{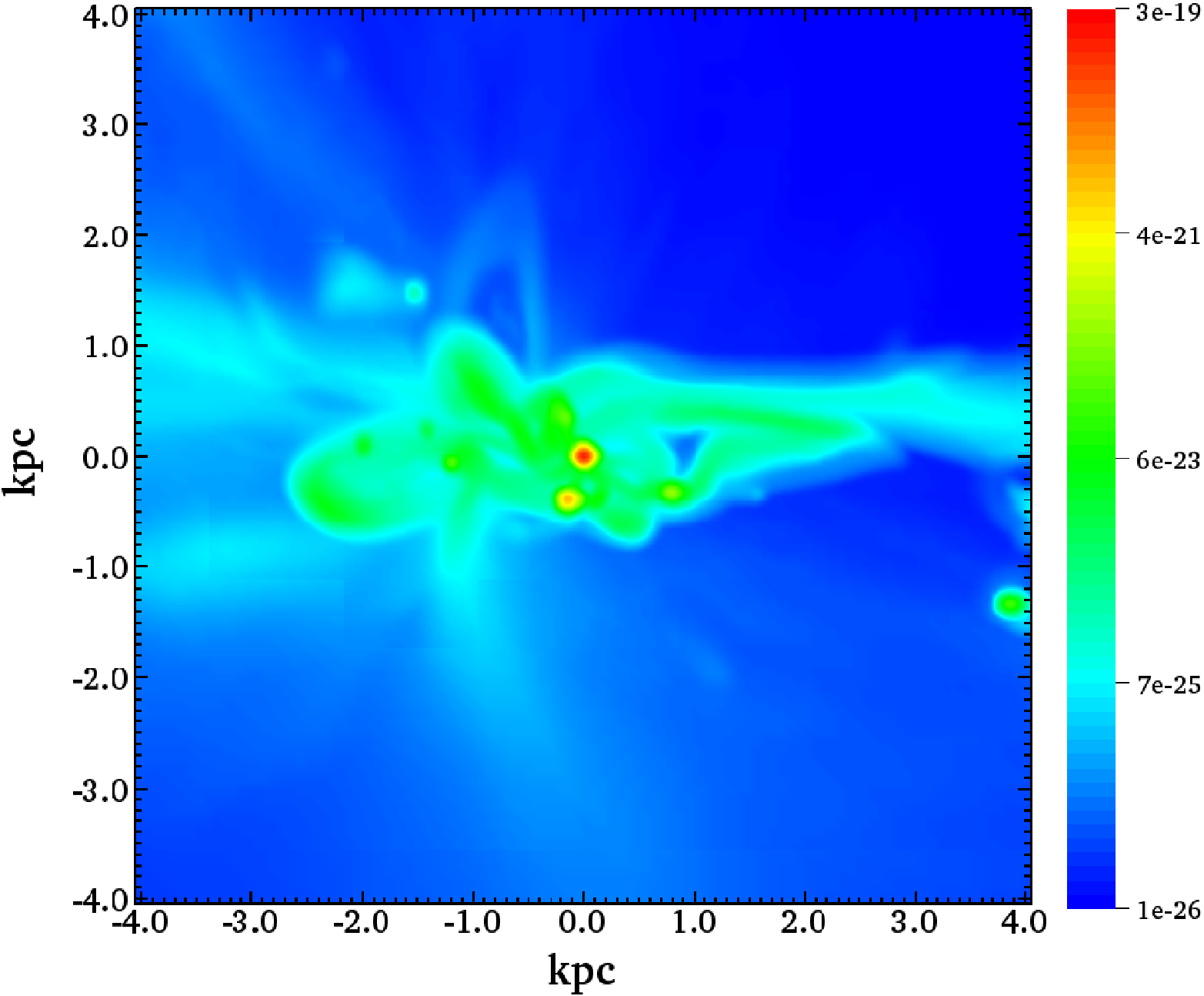}
\end{minipage} &

\begin{minipage}{8cm}
\includegraphics[scale=0.32]{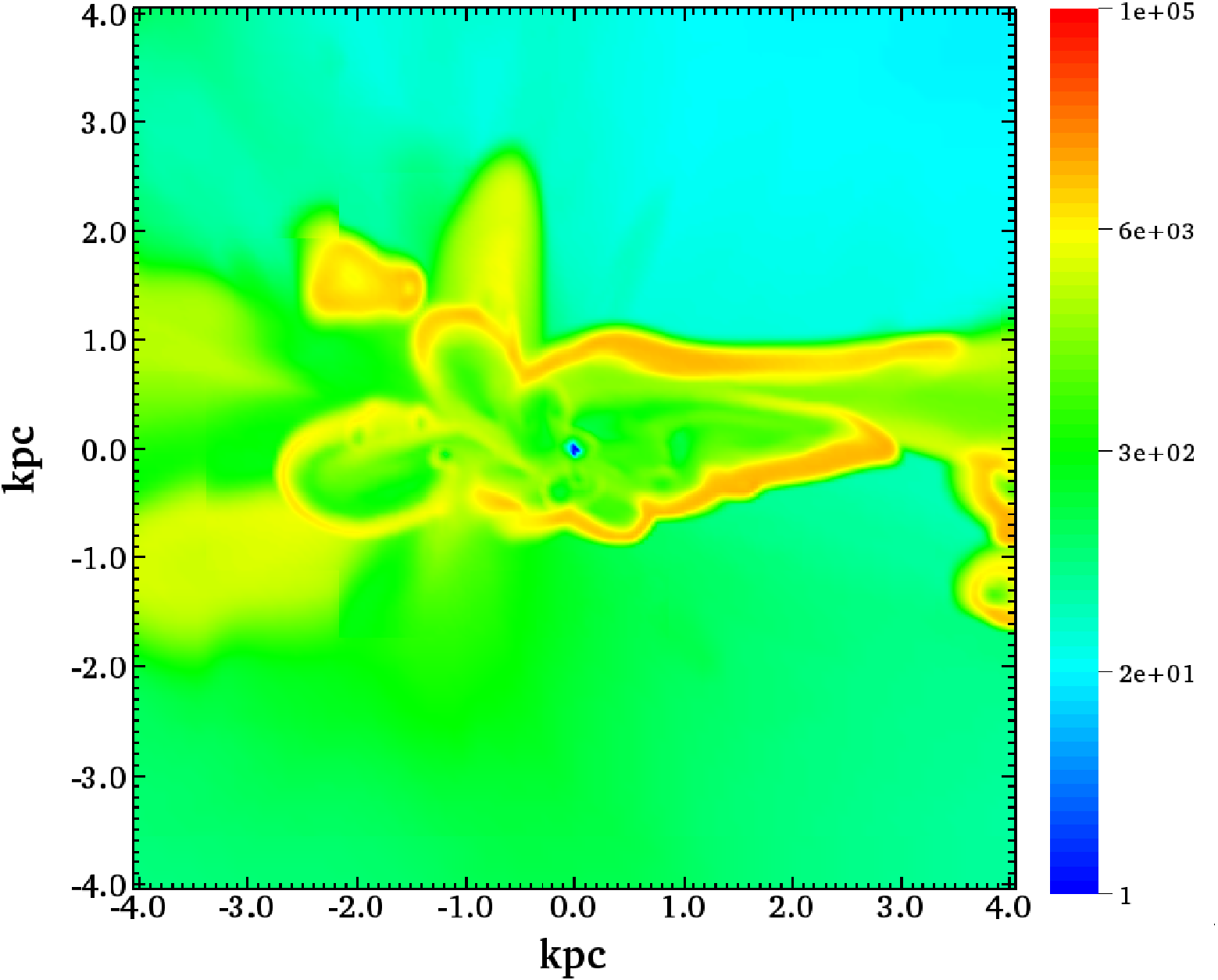}
\end{minipage} \\

\begin{minipage}{8cm}
 \includegraphics[scale=0.28]{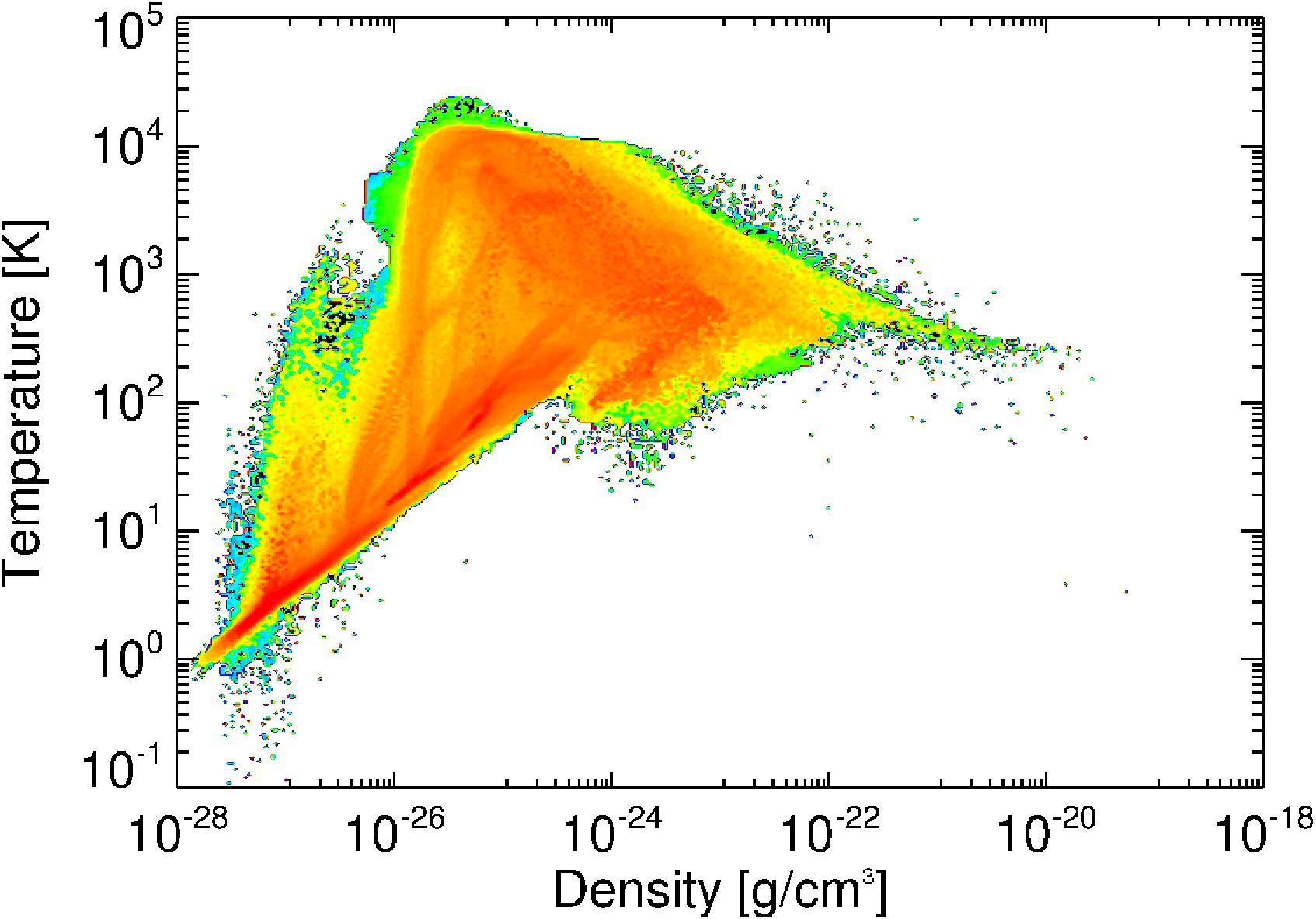}
\end{minipage}&

\begin{minipage}{8cm}
\includegraphics[scale=0.32]{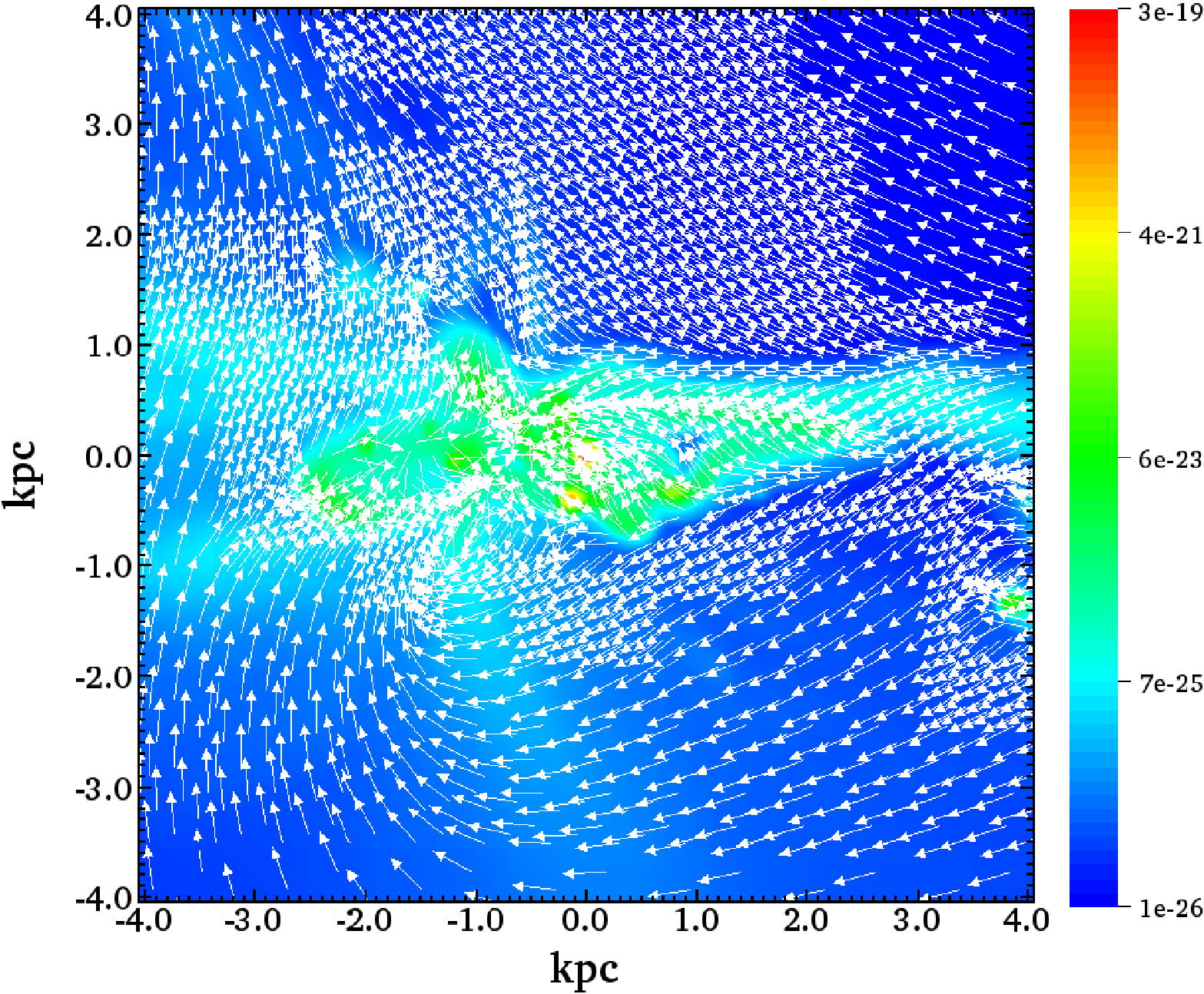}
\end{minipage}

\end{tabular}
\caption{ The figure represents a higher density peak halo for the molecular hydrogen cooling case. The upper left panel shows the density slice through the center of a halo. The temperature slice corresponding to the density slice is depicted in the upper right panel. The bottom left panel shows the phase diagram. The flow of gas is seen in the bottom right panel. Values corresponding to colors are shown in color the bars. Distance scales are in comoving units. Here 1kpc in comoving units corresponds to 53 pc in proper units.}
\label{fig:h2new}
\end{figure*}

\begin{figure*}
\centering
\begin{tabular}{c c}

\begin{minipage}{8cm}
\includegraphics[scale=0.32]{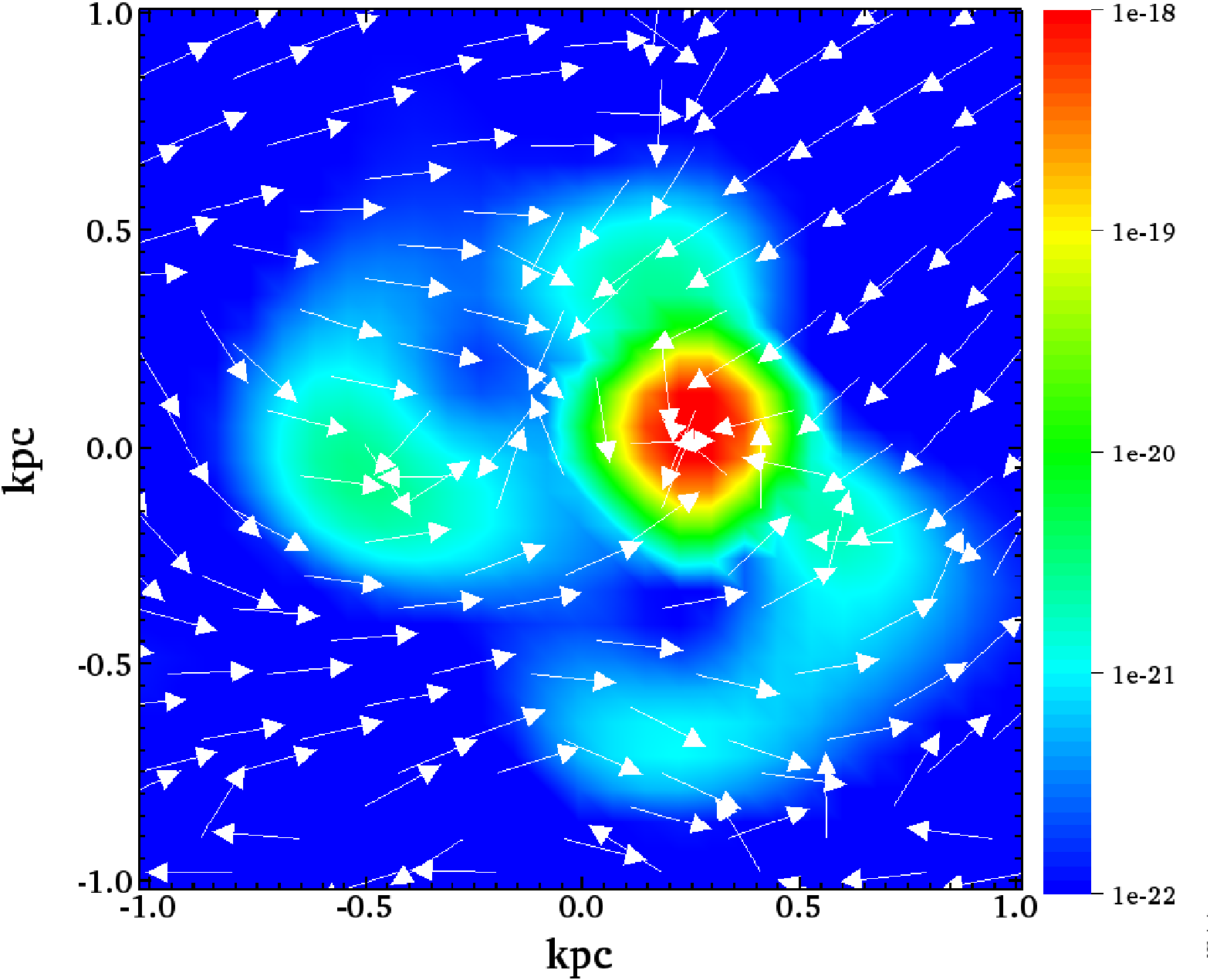}
\end{minipage} &

\begin{minipage}{8cm}
\includegraphics[scale=0.32]{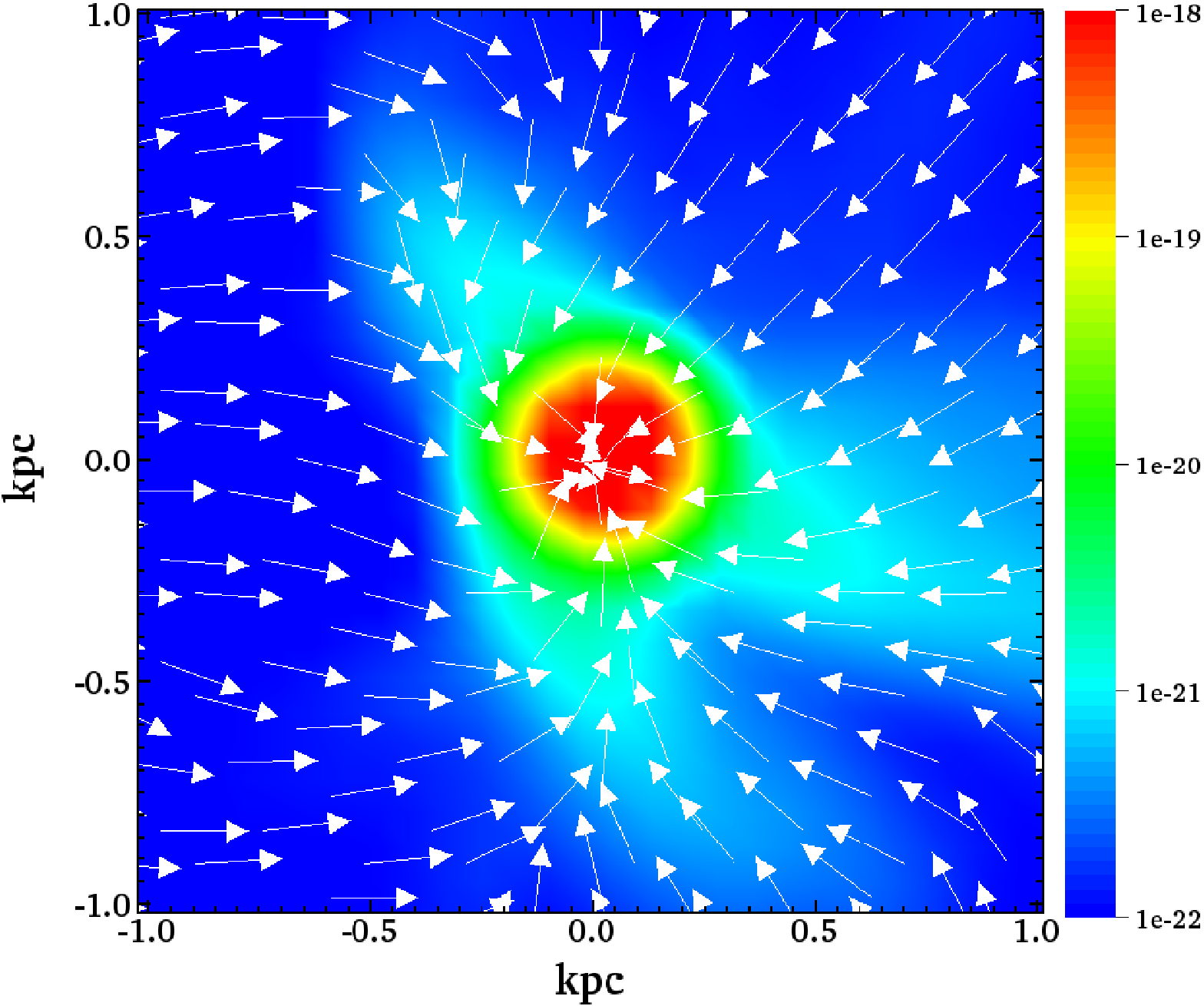}
\end{minipage} \\

\end{tabular}
\caption{Velocity field corresponding to \ref{fig:new3}. The left panel shows the velocity field for the atomic cooling case. The right panel shows velocity field for the Lyman alpha case. Velocity vectors represent the flow of gas after subtracting back ground flow.}
\label{fig:new3-1}
\end{figure*}
The morphology of the object is shown in figure \ref{fig:lymandens}. Diffuse hot gas is visible in the vicinity of the object. Contraction of gas is halted and the Jeans mass is boosted due to Lyman alpha trapping. Diffuse gas may feed the central object during later stages of collapse. The halo does not collapse to high densities (few $\rm \times 10^{3} cm^{-3}$ in this case) as compared to other cases because of suppressed cooling. The temperature is enhanced due to trapping of Lyman alpha photons as depicted in figure \ref{fig:lymantmp}. It is $\rm \geq 10^{4}$K in the center of the halo and even higher in the shock fronts. Trapping of Lyman alpha photons keeps the gas $\rm H_{2}$ free, as collisional dissociation and charge exchange at a temperature above 5000 K destroys $\rm H_{2}$ formation \citep{2002ApJ...569..558O}. \cite{2006ApJ...652..902S} have confirmed through models that column densities in the range of $\rm 10^{22}-10^{25} cm^{-2}$ will be sufficient to self shield the gas even from X-rays. This puts our arguments in favor of suppressed fragmentation on sounder footing. The inability of gas to fragment makes such halos prone to the formation of massive objects. The collapse is never halted but only delayed in the case of Lyman alpha trapping as compared to other simulations (molecular hydrogen and atomic cooling). This delay is caused by the enhanced radiation pressure as more work needs to be done on the gas. Entropy of gas is increased as a consequence of Lyman alpha trapping up to $\rm 0.1~keV cm^{2}$. It is even higher ($\rm 10~keV cm^{2}$) in shocks. We see that trapping of Lyman alpha photons increases the entropy of gas which is close to the lower limit for entropy in simulations of galaxy groups and clusters with radiative cooling \citep{2005MNRAS.361..233B}. We do not include feedback effects which may inject extra energy and raise the entropy of the gas to even higher values.


\section{Discussion and Conclusions}

We have probed the influence of thermodynamics on the formation of the first objects in primordial gas. We have introduced the impact of Lyman alpha trapping, atomic and molecular hydrogen cooling, and performed cosmological three-dimensional simulations down to a redshift of 8.4 using the AMR code FLASH. We exploited the AMR technique to add dynamic levels of refinement wherever required, so that our resolution is sufficient to resolve the most essential physics. We have assumed that halos are metal-free. Pockets of metal-free gas can exist until redshift $\rm \sim$ 6 \citep{2009ApJ...700.1672T}.

Line trapping of Lyman alpha photons stiffens the EOS and enhances the Jeans mass. We noticed that Lyman alpha photons trapped in halos inhibit cooling below $\rm \sim 10^{4}$K, consequently, gas fragmentation is halted and a massive object is formed at the center of a halo. We performed a set of simulation with one higher level of refinement for this case and find similar results as mentioned before. As trapping continues, formation of $\rm H_{2}$ will be suppressed and a halo will collapse adiabatically. The outcome of the object depends on how fast and efficient it accretes mass. It can directly collapse into a black hole or a supermassive star as an interim stage. We conclude that trapping of Lyman alpha photons during gas cloud collapse provides a venue for black holes formation, particularly when there is modest background UV flux to suppress molecular hydrogen cooling. Radiative transfer effects thus play a crucial role in the formation of massive objects observed at $\rm z >6$.

Our results are only valid for metal free halos. The addition of trace amounts of metals can change the scenario for Lyman alpha trapping. If dust is present then the gas temperature is lowered, and the equation of state softened, due to absorption and IR re-emission of Lyman alpha photons. Halos might have been pre-enriched by the heavy metals formed in and dispersed through supernovae explosions \citep{2010ApJ...716..510G,2008MNRAS.388...26J}. Such a scenario would suppress the effectiveness of Lyman alpha trapping because dust would efficiently absorb the Lyman alpha photons and re-radiate them in the far-infrared. As such, we again find an upper limit on the importance of Lyman alpha trapping. Still, if the medium is inhomogeneous, which it very likely is, then Lyman alpha may scatter of dense gas condensations without being absorbed by dust and thus remain trapped in a collapsing structure \citep{1999ApJ...518..138H,1991ApJ...370L..85N}. Fragmentation of the gas is inevitable in the presence of dust and metals \citep{2008ApJ...686..801O}. Recent work by \cite{2010ApJ...712L..69S} shows that after initial collapse, cooling may proceed through other atomic states in the presence of Lyman alpha trapping. It can soften the EOS to unity and lead to disk formation, which may or may not fragment to form a top-heavy stellar initial mass function. We have explored the important phase of Lyman alpha trapped halos. Cosmological simulations should be performed to examine the effect of other cooling channels like 2s-1s and 3-2 transitions in halos that are optically thick to Lyman alpha photons.

For the atomic cooling case, we find that the halo collapses into one or two massive clumps. As long as $\rm H_{2}$ cooling is suppressed, gas collapses isothermally and remains hot having temperature around 8000 K. Further fragmentation is unlikely. Therefore, such halos are plausible candidates for the formation of intermediate mass black holes or supermassive stars as an intermediate stage. The formation of molecular hydrogen does not take place if internal or external UV radiation from the star forming halo can suppress $\rm H_{2}$ formation \citep{2010MNRAS.402.1249S, 2008MNRAS.391.1961D}. At densities $\rm >10^{8} cm^{-3}$ molecular hydrogen may form due to 3-body reactions \citep{1983ApJ...271..632P} in the absence of intense UV flux and may induce fragmentation. The further fate of these objects should be explored through cosmological simulations by including detailed physics.

We have also performed simulations by selecting a high density peak halo. We noticed that in the higher density peak halo the free-fall time becomes shorter and a halo collapses about redshift 25. We followed the collapse down to redshift 18 and compared the results of atomic line cooling with Lyman alpha trapping case. We find that the Lyman alpha trapped halo does not fragment irrespective of the high or low density peak in contrast to the atomic cooling case. Trapping of Lyman alpha photons keeps the gas hot (T$\geq 10^{4}$K) and contraction of a halo is halted. The Jeans mass becomes higher and it does not fragment. The size of the fragment is larger than the atomic cooling case. The mass in a higher density peak halo is $\rm 5\times10^{7}M_{\odot}$. The mass of the object is not much higher than the atomic cooling case because more gas falls into the halo due to efficient atomic cooling and condenses. The density, temperature, phase diagram and velocity vectors for a higher density peak halo are shown in figure \ref{fig:lymannew}.

For the atomic cooling case, morphology and thermodynamical properties of the clumps are similar in both higher and lower density peak halos. The density, temperature, phase diagram and velocity vectors for a higher density peak halo are shown in figure \ref{fig:atomicnew}. We find that the higher density peak halo fragments as in lower density peak case. It is depicted in density slice (zoomed in) in figure \ref{fig:new3}. Density contours are overplotted on density slice and are compared with Lyman alpha case. One can see that there is more structure in atomic case as compared to trapping case. We found that clump surrounding massive one is gravitationally bound and mass is close to the Jeans mass. We also noticed that smaller clump is rotating around massive one. The velocity field for the atomic cooling case (inner 2 kpc region) is shown in figure \ref{fig:new3-1} and is compared with Lyman alpha case. There is also divergence in the flow of gas corresponding to density contours. It supports our claim of fragmentation. The masses of the clumps in higher density peak halo are $\rm \sim 3\times10^{7}M_{\odot},4.0\times10^{5}M_{\odot}$. These fragments will become even more prominent in higher resolution simulations.


In the case of molecular hydrogen cooling, the gas is cooled down to 100 K and collapses into minihalos. The minihalos formed via $\rm H_{2}$ cooling are the potential sites for Pop III star formation. Following the collapse becomes a computationally difficult task as the Jeans mass keeps dropping. We stop the simulation when the Truelove criterion is about to be violated. Adding higher resolution may help in fragmentation. Further fragmentation may occur at the higher densities due to an enhanced fraction of the molecular hydrogen formed through 3-body reactions. We did not include the HD cooling which is a very efficient coolant around 100K. Introduction of the HD cooling will cool the gas below 100K and may enhance fragmentation.

We have performed a set of simulations for the higher density peak $\rm H_{2}$ cooled halo and compared it with the lower density peak halo. We find that the gas collapses into minihalos. Efficient fragmentation of $\rm H_{2}$ cooled gas halo occurs regardless of being in low or high density peak halos. The thermodynamical properties of the minihalos are very similar in both cases. The difference in the collapse time of the higher and lower density peak halos is the same as mentioned for the atomic cooling case. The density, temperature, phase diagram and velocity vectors for a higher density peak halo are shown in figure \ref{fig:h2new}. The masses of the minihalos in this case are $\rm 9\times10^{5}M_{\odot},6\times10^{5}M_{\odot},5\times10^{5}M_{\odot}, 3\times10^{5}M_{\odot},2.6\times10^{4}M_{\odot},~and~ 9\times10^{4}M_{\odot}$. Stars are most likely to form in the $\rm H_{2}$ cooled minihalos. The sizes of the clumps are bigger than $\rm H_{2}$ cooled fragments due to difference in the Jeans masses. Fragmentation to lower than the Jeans scale is not possible as gas becomes pressure supported.

We also see that the entropy of the gas is increased up to $\rm 0.1~keV cm^{2}$ by trapping Lyman alpha photons. It is increased in shocks and lost due to cooling. Complete entropy calculations are beyond the scope of this article. We have not included the feedback effects from the first objects in our simulations. The presence of UV flux from Pop III stars will suppress the formation of $\rm H_{2}$ \citep{2009arXiv0902.2999D,2010MNRAS.402.1249S}. Further fragmentation will be inhibited in these halos. It will support the mode of massive objects formation in the halos that we consider here. It should be noted in this that star formation is not completely suppressed by a UV background due to efficient self-shielding or positive feedback that results from shocks \citep{2007ApJ...665...85J,2002ApJ...575...49R}. X-rays from miniquasars can stimulate the formation of molecular hydrogen \citep{2000ApJ...534...11H}. High columns present in our case will partly self shield the gas from X-rays feedback but the gas in the surrounding might be effected. Removal or redistribution of angular momentum is an obstacle in the formation of massive objects. The gas can transfer angular momentum via shock waves and bar instabilities \citep{2008ApJ...682..745W, 2009arXiv0901.4325L}. We noticed that gas becomes turbulent during virialization and helps in the dissipation of angular momentum. We do see strong shock waves occurring at low densities in our simulations which also play important role in the distribution of angular momentum. Numerical simulations should be performed to get detailed insight about angular momentum issue.

In any case, thermodynamical properties of the gas play an indispensable role in the fragmentation of gas cloud and the formation of primordial objects. Our results include radiative transfer effects of the Lyman alpha photons in an approximate manner, but still give good insight into radiative transfer effects on primordial gas physics. Forthcoming telescopes like LOFAR and JWST will be able to observe the objects formed in the early Universe and will help to scrutinize the present paradigm of primordial objects formation.

\section*{Acknowledgments}

The FLASH code was in part developed by the DOE-supported Alliance Center for Astrophysical Thermonuclear Flashes (ACS) at the University of Chicago. Our simulations were carried out on the Gemini machines at the Kapteyn Astronomical Institute, University of Groningen. We would like to pay special thanks to Seyit Hocuk for valuable discussions. We thank the anonymous referee for careful insight and useful comments.

\label{lastpage}
 
\bibliography{biblio.bib}

\end{document}